\documentclass[12pt]{article}
\usepackage{graphicx}

 \newcommand{\ra}{\rightarrow}

 \newcommand{\equ}{\; \equiv \;}
 \newcommand{\xbf}{\mbox{\boldmath $x$}}

 \newcommand{\imp}{\mbox{\boldmath $p$}}
 
 \newcommand{\kbf}{\mbox{\boldmath $k$}}

 \newcommand{\nabf}{\mbox{\boldmath $\nabla$}}

 \newcommand{\orm}{{\rm o}}
 \newcommand{\pa}{\partial}

 \newcommand{\text}{\rm}
 
 \newcommand{\drm}{{\rm d}}
 \newcommand{\grm}{{\rm g}}

 \newcommand{\Ccal}{{\cal C}}
 \newcommand{\ug}{ \; = \; }

 \newcommand{\ga}{\gamma}

 \newcommand{\infi}{\infty}

 \newcommand{\al}{\alpha}
 \newcommand{\ze}{\zeta}

 \newcommand{\bb}{\begin{equation}}
 \newcommand{\ee}{\end{equation}}
 \newcommand{\bc}{\begin{center}}
 \newcommand{\ec}{\end{center}}
 \newcommand{\bega}{\begin{eqnarray}}
 \newcommand{\ega}{\end{eqnarray}}
 \newcommand{\begae}{\begin{eqnarray*}}
 \newcommand{\egae}{\end{eqnarray*}}

 \newcommand{\h}{\hspace*{4ex}}
 \newcommand{\dis}{\displaystyle}

 \newcommand{\rr}{\rho}

 \newcommand{\Om}{\Omega}
 \newcommand{\om}{\omega}

 \newcommand{\cent}{\centerline}
 \newcommand{\vs}{\vspace*}

\begin{document}
\baselineskip 0.6cm

\begin{center}

{\large {\bf LOCALIZED WAVES: A HISTORICAL AND SCIENTIFIC
INTRODUCTION}$^{\: (\dag)}$} \footnotetext{$^{\: (\dag)}$  Work
partially supported by FAPESP (Brazil), and by MIUR and INFN
(Italy). \ E-mail addresses: \ mzamboni@dmo.fee.unicamp.br; \
recami@mi.infn.it; \ hugo@dmo.fee.unicamp.br}

\end{center}

\vs{5mm}

\cent{ Erasmo Recami }

\vs{0.2 cm}

\cent{{\em Facolt\`a di Ingegneria, Universit\`a statale di
Bergamo, Bergamo, Italy;}} \cent{{\rm and} {\em
INFN---Sezione di Milano, Milan, Italy.}}

\vs{0.5 cm}

\cent{ Michel Zamboni-Rached, }

\vs{0.2 cm}

\centerline{{\em Centro de Ci\^encias Naturais e Humanas,
Universidade Federal do ABC, Santo Andr\'e, SP, Brasil}}

\vs{0.5 cm}

\centerline{\rm and}

\vs{0.3 cm}

\cent{ H. E. Hern\'{a}ndez-Figueroa }

\vs{0.2 cm}

\cent{{\em DMO--FEEC, State University at Campinas,} {\em
Campinas, SP, Brazil.}}

\vs{0.5 cm}

{\bf Abstract  \ --} \ In the first part of this paper (mainly
a review) we present
general and formal (simple) introductions to the ordinary gaussian
waves and to the Bessel waves, by explicitly separating the cases
of the beams from the cases of the pulses; and, finally, an
analogous introduction is presented for the Localized Waves (LW),
pulses or beams.  Always we stress the very different
characteristics of the gaussian with respect to the Bessel waves
and to the LWs, showing the numerous and important properties of
the latter w.r.t. the former ones:  Properties that may find
application in all fields in which an essential role is played by
a wave-equation (like electromagnetism, optics, acoustics,
seismology, geophysics, gravitation, elementary particle physics,
etc.). \h In the second part of this paper (namely, in its
Appendix) we recall at first how, in the seventies and eighties,
the geometrical methods of Special Relativity (SR)  predicted
---in the sense below specified--- the existence of the most
interesting LWs, i.e., of the X-shaped pulses. At last, in
connection with the circumstance that the X-shaped waves are
endowed with Superluminal group-velocities (as carefully discussed
in the first part of this article), we briefly mention the various
experimental sectors of physics in which Superluminal motions seem
to appear: In particular, a bird's-eye view is presented of the
experiments till now performed with evanescent waves (and/or
tunneling photons), and with the ``localized Superluminal
solutions" to the wave equations.

\

\

\section{A GENERAL INTRODUCTION}

\vs{3mm}

\subsection{Preliminary remarks}

Diffraction and dispersion are known since long to be phenomena
limiting the applications of (optical, for instance) beams or
pulses.

\h Diffraction is always present, affecting any waves that
propagate in two or three-dimensional media, even when
homogeneous.  Pulses and beams are constituted by waves traveling
along different directions, which produces a gradual {\em spatial}
broadening\cite{Born}.  This effect is really a limiting factor
whenever a pulse is needed which maintains its transverse
localization, like, e.g., in free space
communications\cite{Willebrand}, image forming\cite{Goodman},
optical lithography\cite{Okazaki,Ito}, electromagnetic
tweezers\cite{Ashkin,Curtis}, etcetera.

\h Dispersion acts on pulses propagating in material media,
causing mainly a temporal broadening: An effect known to be due to
the variation of the refraction index with the frequency, so that
each spectral component of the pulse possesses a different
phase-velocity. This entails a gradual temporal widening, which
constitutes a limiting factor when a pulse is needed which
maintains its {\em time} width, like, e.g., in communication
systems\cite{Agrawal}.

\h It is important, therefore, to develop any techniques able to
reduce those phenomena. \ The so-called {\em localized waves\/}
(LW), known also as non-diffracting waves, are indeed able to
resist diffraction for a long distance in free space. \ Such
solutions to the wave equations (and, in particular, to the
Maxwell equations, under weak hypotheses) were theoretically
predicted long time ago\cite{BarutMR,Stratton,Courant,Bateman}
(cf. also\cite{SupCh}, and the Appendix of this paper),
mathematically constructed in more recent
times\cite{Lu1,PhysicaA}, and soon after experimentally
produced\cite{Lu2,Saari97,Ranfagni}. \ Today, localized waves are
well-established both theoretically and experimentally, and are
having innovative applications not only in vacuum, but also in
material (linear or non-linear) media, showing to be able to
resist also dispersion. \ As we were mentioning, their potential
applications are being intensively explored, always with
surprising results, in fields like Acoustics, Microwaves, Optics,
and are promising also in Mechanics, Geophysics, and even
Gravitational Waves and Elementary particle physics. \ Worth
noticing appear also the applications of the so-called ``Frozen
Waves", that will be presented elsewhere in this book; while
rather interesting are the applications {\em already} obtained,
for instance, in high-resolution ultra-sound scanning of moving
organs in human body\cite{LuBiomedical,LuImaging}.

\h To confine ourselves to electromagnetism, let us recall the
present-day studies on electromagnetic
tweezers\cite{Garces,McGloin,MacDonald,Arlt}, optical (or
acoustic) scalpels, optical guiding of atoms or (charged or
neutral) corpuscles\cite{Rhodes,Fan,Arlt2}, optical
litography\cite{Erdelyi,Garces}, optical (or acoustic)
images\cite{Herman}, communications in free
space\cite{Ziolk,Ziolk2,Lu1,Lu}, remote optical
alignment\cite{Vasara}, optical acceleration of charged
corpuscles, and so on.

\h In the following two Subsections we are going to set forth a
brief introduction to the theory and applications of localized
beams and localized pulses, respectively.

\

{\em Localized (non-diffracting) {\bf beams}} --- The word {\em
beam} refers to a monochromatic solution to the considered wave
equation, with a transverse localization of its field.  To fix our
ideas, we shall explicitly refer to the optical case: But our
considerations, of course, hold for any wave equation (vectorial,
spinorial, scalar...: in particular, for the acoustic case too).

\h The most common type of optical beam is the gaussian one, whose
transverse behavior is described by a gaussian function. But all
the common beams suffer a diffraction, which spoils the transverse
shape of their field, widening it gradually during propagation. As
an example, the transverse width of a gaussian beam doubles when
it travels a distance $z_{\rm
dif}=\sqrt{3}\pi\Delta\rho_0^2/\lambda_0$, where $\Delta\rho_0$ is
the beam initial width and $\lambda_0$ is its wavelength. One can
verify that a gaussian beam with an initial transverse aperture of
the order of its wavelength will already double its width after
having travelled along a few wavelenths.

\h It was generally believed that the only wave devoid of
diffraction was the plane wave, which does not suffer any
transverse changes. Some authors had shown, actually, that it
isn't the only one. For instance, in 1941 Stratton\cite{Stratton}
obtained a monochromatic solution to the wave equation whose
transverse shape was concentrated in the vicinity of its
propagation axis and represented by a Bessel function. Such a
solution, now called a Bessel beam, was not subject to
diffraction, since no change in its transverse shape took place
with time. In ref.\cite{Courant} it was later on demonstrated how
a large class of equations (including the wave equations) admit
``non-distorted progressing waves" as solutions; while already in
1915, in ref.\cite{Bateman}, and subsequently in articles like
ref.\cite{Barut2}, it was shown the existence of soliton-like,
wavelet-type solutions to the Maxwell equations. But all such
literature did not raise the attention it deserved. In the case of
ref.\cite{Stratton}, this can be partially justified since that
(Bessel) beam was endowed with infinite energy [as much as the
plane waves].  An interesting problem, therefore, was that of
investigating what it would happen to the ideal Bessel beam
solution when truncated by a finite transverse aperture.

\h Only in 1987 a heuristical answer came from the known
experiment by Durnin et al.\cite{Durnin}, when it was shown that a
realistic Bessel beam, endowed with wavelength
$\lambda_0=0.6328\;\mu$m and central spot\footnote{Let us define
the central ``spot'' of a Bessel beam as the distance, along the
propagation axis $\rho=0$, at which the first zero occurs of the
Bessel function characterizing its transverse shape.}
$\Delta\rho_0=59\;\mu$m, passing through an aperture with radius
$R=3.5\;$mm is able to travel about $85\;$cm keeping its
transverse intensity shape approximately unchanged (in the region
$\rho << R$ surrounding its central peak). In other words, it was
experimentally shown that the transverse intensity peak, as well
as the field in the surroundings of it, do not meet any appreciable
change in shape all along a {\em large} ``depth of field". As a
comparison, let us recall once more that a gaussian beam with the
same wavelength, and with the central ``spot''\footnote{In the
case of a gaussian beam, let us define its central ``spot'' as the
distance along $\rho=0$ at which its transverse intensity has
decayed of the factor $1/e$.} $\Delta\rho_0=59\;\mu$m, when
passing through an aperture with the same radius $R=3.5\;$mm,
doubles its transverse width after $3\;$cm, and after $6\;$cm its
intensity is already diminished of a factor 10. \ Therefore, in
the considered case, a Bessel beams can travel, approximately
without deformation, a distance 28 times larger than a gaussian
beam's.

\h Such a remarkable property is due to the fact that the
transverse intensity fields (whose value decreases with increasing
$\rho$), associated with the rings which constitute the
(transverse) structure of the Bessel beam, when diffracting, end
up {\em reconstructing} the beam itself, all along a large
field-depth. All this depends on the Bessel beam spectrum
(wavenumber and frequency),\cite{Herman,Durnin2,Vasara} as
explained in detail in our ref.\cite{MRH}. \ Let us stress that,
given a Bessel and a gaussian beam ---both with the same spot
$\Delta\rho_0$ and passing through apertures with the same radius
$R$ in the plane $z=0$, and with the same energy $E$--- the
percentage of the total energy $E$ contained inside the central
peak region ($0 \leq \rho \leq \Delta\rho_0$) is smaller for a
Bessel than for a gaussian one: This different energy-distribution
on the transverse plane is responsible for the reconstruction of
the Bessel-beam central peak even at large distances from the
source (and even after an obstacle with a size smaller than the
aperture\cite{Bouchal,Michel,Grunwald}: a nice property possessed
also by the localized pulses we are going to examine
below\cite{Michel}).

\begin{figure}[!h]
\begin{center}
 \scalebox{1.6}{\includegraphics{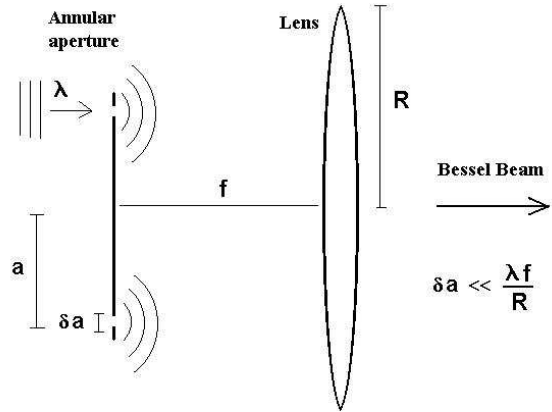}}  
\end{center}
\caption{The simple experimental set-up used by Durnin et al. for
generating a Bessel beam.} \label{fig1}
\end{figure}

\h It may be worth mentioning that most experiments carried on in
this area have been performed rapidly and with use, often, of
rather simple apparata: The Durnin et al.'s experiment, e.g., had
recourse, for the generation of a Bessel beam, to a laser source,
an annular slit and a lens, as depicted in Fig.(\ref{fig1}). In a
sense, such an apparatus produces what can be regarded as the
cylindrically symmetric generalization of a couple of plane waves
emitted at angles $theta$ and $- theta$, w.r.t. the $z$-direction,
respectively (in which case the plane wave {\em intersection}
moves along $z$ with the speed $c/\sin \theta$). \ Of course,
these non-diffracting beams can be generated also by a a conic
lens ({\em axicon\/}) [cf., e.g., ref.\cite{Herman}], or by other
means like holographic elements [cf., e.g.,
refs.\cite{Vasara,MacDonald2}].

\h Let us emphasize, as already mentioned at the end of the
previous Subsection, that nowadays a lot of interesting
applications of non-diffracting beams are being investigated;
besides the Lu et al.'s ones in Acoustics. In the optical sector,
let us recall again those of using Bessel beams as optical
tweezers able to confine or move around small particles. \ In such
theoretical and application areas, a noticeable contribution is
the one presented in refs.\cite{MichelOE1,FWart2,MichelOE2},
wherein, by suitable superpositions of Bessel beams endowed with
the same frequency but different longitudinal wavenumbers, {\em
stationary} fields have been mathematically constructed in closed
form, which possess a high transverse localization and, more
important, a longitudinal intensity-shape that can be freely
chosen inside a predetermined space-interval $0 \leq z \leq L$.
For instance, a high intensity field, with a static envelope, can
be created within a tiny region, with negligible intensity
elsewhere: Chapter 2 of the coming book [{\em Localized Waves} (J.Wiley; in
press] will deal, among the others, with such ``Frozen Waves".

\

{\em Localized (non-diffracting) {\bf pulses}} --- As we have seen
in the previous Subsection, the existence of non-diffractive (or
localized) {\em pulses} was predicted since long: cf., once more,
refs.\cite{Bateman,Courant}, and, not less,
refs.\cite{BarutMR,SupCh}, as well as more recent articles like
refs.\cite{Barut3,Barut4}. \ The modern studies about
non-diffractive pulses (to confine ourselves, at least, to the
ones that attracted more attention) followed a development rather
independent of those on non-diffracting {\em beams}, even if both
phenomena are part of the same sector of physics: that of {\em
Localized Waves}.

\h In 1983, Brittingham\cite{Brittingham} set forth a luminal
($V=c$) solution to the wave equation (more particularly, to the
Maxwell equations) with travels rigidly, i.e., without
diffraction. The solution proposed in ref.\cite{Brittingham}
possessed however infinite energy, and once more the problem arose
of overcoming such a problem.

\h A way out was first obtained, as far as we know, by
Sezginer\cite{Sezginer}, who showed how to construct finite-energy
luminal pulses, which ---however--- do not propagate without
distortion for an infinite distance, but, as it is expected,
travel with constant speed, and approximately without deforming,
for a certain ({\em long\/} depth of field: much longer, in this
case too, than that of the ordinary pulses like the gaussian ones.
In a series of subsequent
papers\cite{Ziolk,Ziolk2,BSZ,SBZ,ZBS,Donnelly}, a simple
theoretical method was developed, called by those authors
``bidirectional decomposition", for constructing a new series of
non-diffracting, {\em luminal} pulses.

\begin{figure}[!h]
\begin{center}
 \scalebox{1.1}{\includegraphics{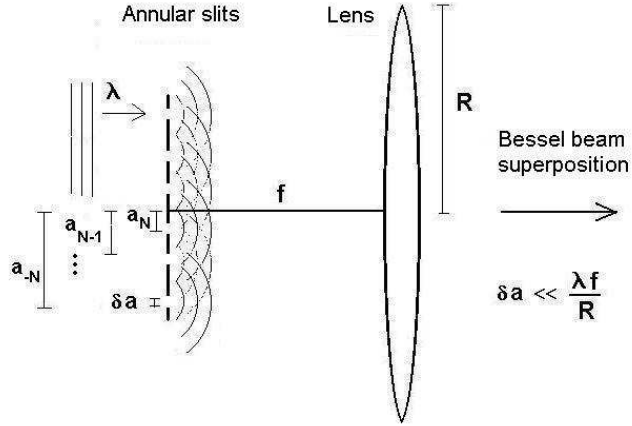}}  
\end{center}
\caption{One of the simplest experimental set-ups for generating
various kinds of Bessel beam superpositions.} \label{fig2}
\end{figure}

\h Eventually, at the beginning of the nineties, Lu et
al.\cite{Lu1,Lu2} constructed, both mathematically and
experimentally, new solutions to the wave equation in free space:
namely, an X-shaped localized pulse, with the form predicted by the
so-called extended Special Relativity\cite{BarutMR,Review}; for
the connection between what Lu et al. called ``X-waves" and
``extended" relativity see, e.g., ref.\cite{SupCh}, while brief
excerpts of that theory can be found, for instance, in
refs.\cite{JSTQE,birdseye,PhysicaA,RFG,FoP87}. \ Lu et al.'s
solutions were continuous superpositions of Bessel beams with the
same phase-velocity (i.e., with the same axicon
angle\cite{Friberg,PhysicaA,Lu1,Review}, $alpha$): cf., e.g.,
Fig.(\ref{fig2}); so that they could keep their shape for long
distances. \ Such X-shaped waves resulted to be interesting and
flexible localized solutions, and have been afterwards studied in
a number of papers, even if their velocity $V$ is supersonic or
Superluminal ($V>c$): Actually, when the phase-velocity does not
depend on the frequency, it is known that such a phase-velocity
becomes the group-velocity! \ Remembering how a superposition of
Bessel beams is generated (for example, by a discrete or
continuous set of annular slits or transducers), it results clear
that the energy forming the X-waves, coming from those rings,
travels at the ordinary speed $c$ of the plane waves in the
considered medium\cite{ZBS2,SB,PhysicaA,SB2}  [here $c$,
representing the velocity of the plane waves in the medium, is the
sound-speed in the acoustic case, and the speed of light in the
electromagnetic case; and so on]. \ {\em Nevertheless}, the peak
of the X-shaped waves is {\em faster} than $c$.

\begin{figure}[!h]
\begin{center}
 \scalebox{3}{\includegraphics{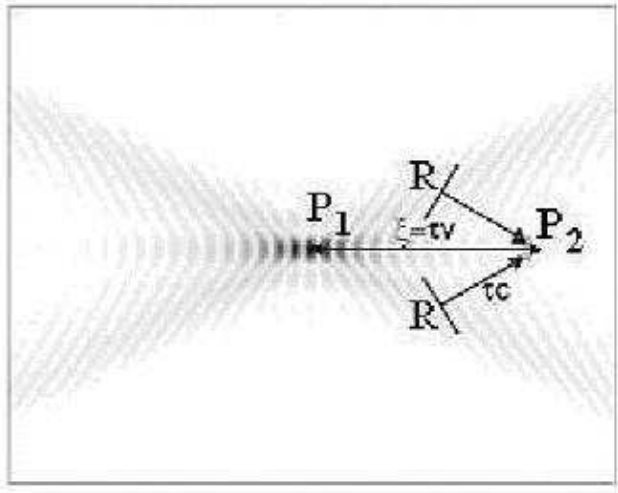}}        
\end{center}
\caption{This figure shows an X-shaped wave, that is, a localized
Superluminal pulse.  It refers to an X-wave, possessing the
velocity $V>c$, and illustrates the fact that, if its vertex or
central spot is located at $P_1$ at time $t_0$, it will reach the
position $P_2$ at a time $t+\tau$ where $\tau=|P_2-P_1|/V \; < \;
|P_2-P_1|/c$: This is something different from the illusory
``scissor effect", {\em even if the feeding energy, coming from
the regions $R$, has travelled with the ordinary speed $c$} (which
is the speed of light in the electromagnetic case, or the sound
speed in Acoustics, and so on).} \label{fig3}
\end{figure}

\h It is possible to generate (besides the ``classic" X-wave
produced by Lu et al. in 1992) infinite sets of new X-shaped
waves, with their energy more and more concentrated in a spot
corresponding to the vertex region\cite{MRH}. It may therefore
appear rather intriguing that such a spot [even if no violations
of special relativity (SR) is obviously implied: all the results
come from Maxwell equations, or from the wave
equations\cite{Barbero,Brodowsky}]--- travels Superluminally when
the waves are electromagnetic. \ We shall call ``Superluminal" all
the X-shaped waves, even, e.g., when the waves are acoustic. \ By
Fig.(\ref{fig3}), which refers to an X-wave possessing the
velocity $V>c$, we illustrate the fact that, if its vertex or
central spot is located at $P_1$ at time $t_1$, it will reach the
position $P_2$ at a time $t+\tau$ where $\tau=|P_2-P_1|/V \; < \;
|P_2-P_1|/c$: \ We shall discuss all these points below.

\h Soon after having mathematically and experimentally constructed
their ``classic" {\em acoustic} X-wave, Lu et al. started applying
them to ultrasonic scanning, obtaining ---as we already said---
very high quality images. \ Subsequently, in a 1996 e-print and
report, Recami et al. (see, e.g., ref.\cite{PhysicaA} and refs.
therein) published the analogous X-shaped solutions to the Maxwell
equations: By constructing scalar Superluminal localized solutions
for each component of the Hertz potential. That showed, by the
way, that the localized solutions to the scalar equation can be
used ---under very weak conditions--- for obtaining localized
solutions to Maxwell's equations too (actually, Ziolkowski et
al.\cite{Ziolkowski0} had found something similar, called by them
{\em slingshot} pulses, for the simple scalar case; but their
solution had gone practically unnoticed). \ In 1997 Saari et
al.\cite{Saari97} announced, in an important paper, the production
in the lab of an X-shaped wave in the optical realm, thus proving
experimentally the existence of Superluminal electromagnetic
pulses. \ Three years later, in 2000, Mugnai et al.\cite{Ranfagni}
produced, in an experiment of theirs, Superluminal X-shaped waves
in the microwave region [their paper aroused various criticisms,
to which those author however responded].

%
%
%
%

\

\section{A MORE DETAILED INTRODUCTION}

\h Let us refer\cite{TesiM} to the differential equation known as
homogeneous wave equation: simple, but so important in Acoustics,
Electromagnetism (Microwaves, Optics,...), Geophysics, and even,
as we said, gravitational waves and elementary particle physics:

\

\bb
 \left(\frac{\pa^2}{\pa x^2} + \frac{\pa^2}{\pa y^2} +
\frac{\pa^2}{\pa z^2} - \frac{1}{c^2}\frac{\pa^2}{\pa
t^2}\right)\, \psi(x,y,z;t) \ug 0 \label{eo1}\ee

\

\h Let us write it in the cylindrical co-ordinates $(\rho,\phi,z)$
and, for simplicity's sake, confine ourselves to axially symmetric
solutions $\psi(\rho,z;t)$. \ Then, eq.(\ref{eo1}) becomes

\

\bb \left(\frac{\pa^2}{\pa\rho^2} +
\frac{1}{\rho}\frac{\pa}{\pa\rho} + \frac{\pa^2}{\pa z^2} -
\frac{1}{c^2}\frac{\pa^2}{\pa t^2} \right)\, \psi(\rho,z;t) \ug 0
\ . \label{eo2}\ee

\

\h In free space, solution $\psi(\rho,z;t)$ can be written in
terms of a Bessel-Fourier transform w.r.t. the variable $\rho$,
and two Fourier transforms w.r.t. variables $z$ and $t$, as
follows:

\

\bb \psi(\rho,z,t) \ug
\int_{0}^{\infi}\int_{-\infi}^{\infi}\int_{-\infi}^{\infi}\,k_\rr
\, J_0(k_\rr\,\rho)\,e^{ik_zz}\,e^{-i\om
t}\,\bar{\psi}(k_\rr,k_z,\om)\,\drm k_\rr\, \drm k_z\,\drm\om
\label{sg1}\ee

\

where $J_0(.)$ is an ordinary zero-order Bessel function and
$\bar{\psi}(k_\rr,k_z,\om)$ is the transform of $\psi(\rho,z,t)$.

\h Substituting eq.(\ref{sg1}) into eq.(\ref{eo2}), one obtains
that the relation, among $\om$, $k_\rr$ and $k_z \,$,

\

\bb \frac{\om^2}{c^2} \ug k_\rr^2 + k_z^2  \label{c1} \ee

\

has to be satisfied. \ In this way, by using condition (\ref{c1})
in eq.(\ref{sg1}), any solution to the wave equation (\ref{eo2})
can be written

\

\bb \psi(\rho,z,t) \ug
\dis{\int_{0}^{\om/c}\int_{-\infi}^{\infi}\, k_\rr \,
J_0(k_\rr\,\rho)\,e^{i\sqrt{\om^2/c^2
\,-\,k_\rr^2}\,\,z}\,e^{-i\om t}\,S(k_\rr,\om)\,\drm
k_\rr\,\drm\om} \label{sg} \ee

\

where $S(k_\rr,\om)$ is the chosen spectral function.

\h The general integral solution (\ref{sg}) yields for instance
the ({\bf non-localized}) gaussian beams and pulses, to which we
shall refer for illustrating the differences of the localized
waves w.r.t. them.

\

{\em The Gaussian Beam} --- A very common (non-localized) beam is
the gaussian beam\cite{Molone}, corresponding to the spectrum

\

\bb S(k_\rr,\om) \ug 2a^2\,e^{-a^2 k_\rr^2}\,\delta(\om - \om_0)
\label{eg} \ee

\

In eq.(\ref{eg}), $a$ is a positive constant, which will be shown
to depend on the transverse aperture of the initial pulse.

\begin{figure}[!h]
\begin{center}
 \scalebox{2.3}{\includegraphics{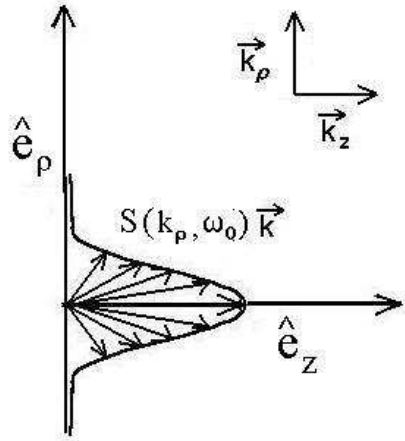}} 
\end{center}
\caption{Visual interpretation of the integral solution
(\ref{sg}), with the spectrum (\ref{eg}), in terms of a
superposition of plane waves.} \label{fig4}
\end{figure}

\h Figure \ref{fig4} illustrates the interpretation of the
integral solution (\ref{sg}), with spectral function (\ref{eg}),
as a superposition of plane waves. \ Namely, from Fig.\ref{fig4}
one can easily realize that this case corresponds to plane waves
propagating in all directions (always with $k_z \geq 0$), the most
intense ones being those directed along (positive) $z$. Notice
that in the plane-wave case $\vec{k_z}$ is the longitudinal
component of the wave-vector, $\vec{k} = \vec{k_\rr} + \vec{k_z}$,
where $\vec{k_\rr} = \vec{k_x}+\vec{k_y}$.

\h On substituting eq.(\ref{eg}) into eq.(\ref{sg}) and adopting
the paraxial approximation, one meets the gaussian beam

\

\bb \psi_{\rm gauss}(\rho,z,t) \ug \frac{\dis{2a^2\,{\rm
exp}\left(\dis{\frac{-\rho^2}{4(a^2 +
i\,z/2k_0)}}\right)}}{\dis{2(a^2 +
i\,z/2k_0)}}\,\,\dis{e^{ik_0(z-ct)}}\;\; , \label{fg}\ee

\

where $k_0=\om_0/c$.  We can verify that such a beam, which
suffers transverse diffraction, doubles its initial width
$\Delta\rho_0 = 2a$ after having traveled the distance $z_{{\rm
dif}} \ug \sqrt{3}\,k_0 \Delta\rho_0^2 /2$, called diffraction
length. The more concentrated a gaussian beam happens to be, the
more rapidly it gets spoiled.

\

{\em The Gaussian Pulse} --- The most common (non-localized) {\em
pulse} is the gaussian pulse, which is got from eq.(\ref{sg}) by
using the spectrum\cite{chirped}

\

\bb S(k_\rr,\om) \ug
\frac{2ba^2}{\sqrt{\pi}}e^{-a^2k_\rr^2}e^{-b^2(\om-\om_0)^2}
\label{epg} \ee

\

where $a$ and $b$ are positive constants.  Indeed, such a pulse is
a superposition of gaussian beams of different frequency.

\h Now, on substituting eq.(\ref{epg}) into eq.(\ref{sg}), and
adopting once more the paraxial approximation, one gets the
gaussian pulse:

\

\bb \psi(\rho,z,t) \ug \frac{a^2\,{\rm
exp}\left(\dis{\frac{-\rho^2}{4(a^2+iz/2k_0)} }\right){\rm
exp}\left(\dis{\frac{-(z-ct)^2}{4c^2b^2)}}
\right)}{a^2+iz/2k_0}\;\;, \label{pg} \ee

\

endowed with speed $c$ and temporal width $\Delta t = 2b$, and
suffering a progressive enlargement of its transverse width, so
that its initial value gets doubled already at position $z_{{\rm
dif}} \ug \sqrt{3}\,k_0 \Delta\rho_0^2/2 \;$, with $\Delta\rho_0 =
2a$.

\

\subsection{The localized solutions}

Let us finally go on to the construction of the two most renowned
localized waves: the Bessel beam, and the ordinary X-shaped
pulse.\cite{TesiM}

\h First of all, it is interesting to observe that, when
superposing (axially symmetrical) solutions of the wave equation
in the vacuum, three spectral parameters come into the play, $(\om, \
k_\rr, \ k_z)$, which have however to satisfy the constraint
(\ref{c1}), deriving from the wave equation itself. Consequently,
only two of them are independent: and we choose\footnote{Elsewhere
we chose $\om$ and $k_z$.} here $\om$ and $k_\rr$. \ Such a
freedom in choosing $\om$ and $k_\rr$ was already apparent in the
spectral functions generating gaussian beams and pulses, which
consisted in the product of two functions, one depending only on
$\om$ and the other on $k_\rr$.

\h We are going to see that particular relations between $\om$ and
$k_\rr$ [or, analogously, between $\om$ and $k_z$] can be moreover
imposed, in order to get interesting and unexpected results, such
as the {\em localized waves}.

\

{\em The Bessel beam} --- Let us start by imposing a {\em linear}
coupling between $\om$ and $k_\rr$ (it could be actually
shown\cite{Durnin2} that it is the unique coupling leading to
localized solutions).

\h Namely, let us consider the spectral function

\

\bb S(k_\rr,\om) \ug \frac{\delta(k_\rr -
\dis{\frac{\om}{c}}\sin\theta)}{k_\rr}\,\,\delta(\om - \om_0)
\;\;, \label{eb} \ee

\

which implies that $k_\rr = (\om\sin\theta)/c$, with $0 \leq
\theta \leq \pi/2$: a relation that can be regarded as a
space-time coupling. Let us add that this linear constraint
between $\om$ and $k_\rr$, together with relation (\ref{c1}),
yields $k_z = (\om\cos\theta)/c$. This is an important fact, since
it has been shown elsewhere\cite{MRH} that an {\em ideal}
localized wave {\em must} contain a coupling of the type $\om=V
k_z + b$, where $V$ and $b$ are arbitrary constants.

\begin{figure}[!h]
\begin{center}
 \scalebox{2}{\includegraphics{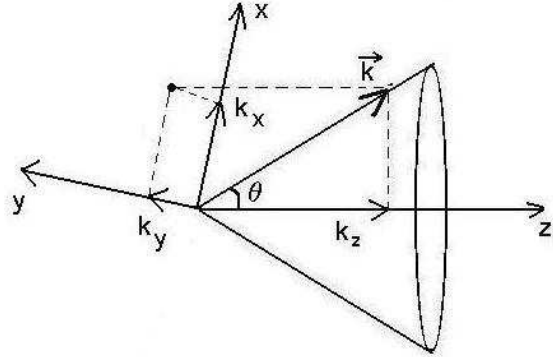}} 
\end{center}
\caption{The axially-symmetric Bessel beam is created by the
superposition of plane waves whose wave vectors lay on the surface
of a cone having the propagation axis as its symmetry axis and
angle equal to $\theta$ ("axicon angle").} \label{fig5}
\end{figure}

\h The interpretation of the integral function (\ref{sg}), this
time with the spectrum (\ref{eb}), as a superposition of plane
waves is visualized by Figure \ref{fig5}: which shows that an
axially-symmetric Bessel beam is nothing but the result of the
superposition of plane waves whose wave vectors lay on the surface
of a cone having the propagation axis as its symmetry axis and an
opening angle equal to $\theta$; such $\theta$ being called the
{\em axicon angle}.

\h By inserting eq.(\ref{eb}) into eq.(\ref{sg}), one gets the
mathematical expression of the so-called Bessel beam:

\

\bb \psi(\rho,z,t) \ug J_0\left(\frac{\om_0}{c}\sin\theta\,\,\rho
\right)\,{\rm
exp}\left[i\,\,\frac{\om_0}{c}\cos\theta\,\left(z-\frac{c}{\cos\theta}t\right)\right]
\label{fb} \ee

\

\h This beam possesses phase-velocity $v_{\rm ph}=c/\cos\theta$,
and field transverse shape represented by a Bessel function
$J_0(.)$ so that its field in concentrated in the surroundings of
the propagation axis $z$. Moreover, eq.(\ref{fb}) tells us that
the Bessel beam keeps its transverse shape (which is therefore
invariant) while propagating, with central ``spot'' $\Delta\rho =
2.405 c /(\om\sin\theta)$.

\h The ideal Bessel beam, however, is not a square-integrable
function, and possesses therefore an infinite energy, i.e., it
cannot be experimentally produced.

\h But we can have recourse to truncated Bessel beams, generated
by finite apertures. In this case the (truncated) Bessel beams are
still able to travel a long distance while maintaining their
transfer shape, as well as their speed, approximately
unchanged\cite{Durnin,Durnin3,Overfelt}: That is to say, they
still possess a large field-depth. For instance, the depth of
field of a Bessel beam generated by a circular finite aperture
with radius $R$ is given by

\

\bb Z_{\rm max} \ug \frac{R}{\tan\theta} \ee

\begin{figure}[!h]
\begin{center}
 \scalebox{2.5}{\includegraphics{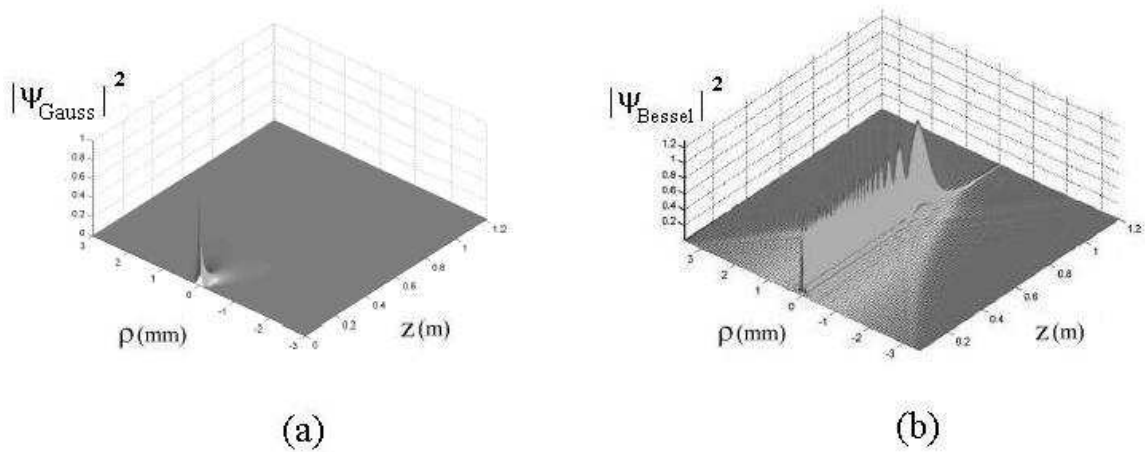}}  
\end{center}
\caption{Comparison between a gaussian (a) and a truncated Bessel
beam (b). One can see that the gaussian beam doubles its initial
transverse width already after $3\;$cm, while after $6\;$cm its
intensity decays of a factor 10.  By contrast, the Bessel beam
does approximately keep its transverse shape till the distance
$85\;$cm.} \label{fig6}
\end{figure}

\

where $\theta$ is the beam axicon angle. In the finite aperture
case, the Bessel beam cannot be represented any longer by
eq.(\ref{fb}), and one has to calculate it by the scalar
diffraction theory: Using, for example, Kirchhoff's or
Rayleigh-Sommerfeld's diffraction integrals. But until the
distance $Z_{\rm max}$ one may still use eq.(\ref{fb}) for
approximately describing the beam, at least in the vicinity of the
axis $\rho=0$; namely, for $\rho << R$. To realize how much a
truncated Bessel beam succeeds in resisting diffraction, let us
consider also a gaussian beam, with the same frequency and central
``spot", and compare their field-depths. In particular, let us
assume for both beams $\lambda = 0.63\,\mu$m and initial central
``spot'' size $\Delta\rho_0 = 60\,\mu$m. The Bessel beam will
possess axicon angle $\theta=\arcsin(2.405
c/(\om\Delta\rho_0))=0.004\;$rad. Figure \ref{fig6} depicts the
behavior of the two beams for a Bessel beam circular aperture with
radius $3.5\;$mm. \ We can see how the gaussian beam doubles its
initial transverse width already after $3\;$cm, and after $6\;$cm
its intensity has become an order of magnitude smaller. By
contrast, the truncated Bessel beam keeps its transverse shape
untill the distance $Z_{\rm max}=R/\tan\theta= 85\;$cm. Afterwards,
the Bessel beam rapidly decays, as a consequence of the sharp cut
performed on its aperture (such cut being responsible also for the
intensity oscillations suffered by the beam along its propagation
axis, and for the fact that eventually the feeding waves, coming
from the aperture, at a certain point get faint).

\h The zeroth-order (axially symmetric) Bessel beam is nothing but
one example of localized beam. Further examples are the higher
order (not cylindrically symmetric) Bessel beams

\bb \psi(\rho,\phi,z;t) \ug
J_{\nu}\left(\frac{\om_0}{c}\sin\theta\,\,\rho \right)\,{\rm
exp}(i\nu\phi) \; {\rm
exp}\left(i\,\,\frac{\om_0}{c}\cos\theta\,\left(z-\frac{c}
{\cos\theta}t\right)\right)\;\; , \label{fbn} \ee

or the Mathieu beams\cite{Dartora1}, and so on.

\

{\em The Ordinary X-shaped Pulse} --- Following the same procedure
adopted in the previous subsection, let us construct pulses by
using spectral functions of the type

\

\bb S(k_\rr,\om) \ug \frac{\delta(k_\rr -
\dis{\frac{\om}{c}}\sin\theta)}{k_\rr}\,\,F(\om) \label{ex} \ee

\

where this time the Dirac delta function furnishes the spectral
space-time coupling $k_\rr = (\om\sin\theta)/c$. Function $F(\om)$
is, of course, the frequency spectrum; it is left for the moment
undetermined.

\h On using eq.(\ref{ex}) into eq.(\ref{sg}), one obtains

\

\bb \psi(\rho,z,t) \ug \int_{-\infty}^{\infty}\,F(\om)\,
J_0\left(\frac{\om}{c}\sin\theta\,\,\rho \right)\,{\rm
exp}\left(\frac{\om}{c}\cos\theta\,\left(z-\frac{c}
{\cos\theta}t\right)\right) \, \drm\om \ . \label{px} \ee

\

It is easy to see that $\psi$ will be a pulse of the type

\

\bb \psi \ug \psi(\rho,z-Vt) \ee

\

with a speed $V=c/\cos\theta$ independent of the frequency
spectrum $F(\om)$.

Such solutions are known as X-shaped pulses, and are {\em
localized} (non-diffractive) waves in the sense that they maintain
their spatial shape during propagation (see., e.g.,
refs.\cite{Lu1,PhysicaA,MRH} and refs. therein).

\

\h {\em At this point, some remarkable observations are in order:}

\

(i) When a pulse consists in a superposition of waves (in this
case, Bessel beams) all endowed with the same phase-velocity
$V_{\rm ph}$ (in this case, with the same axicon angle)
independent of their frequency, {\em then} it is known that the
phase-velocity (in this case $V_{\rm ph}=c/\cos\theta$) becomes
the group-velocity\cite{Majorana,MRF} $V$: That is, $V=
c/\cos\theta >c$. In this sense, the X-shaped waves are called
``Superluminal localized pulses" (cf., e.g., ref.\cite{PhysicaA}
and refs. therein).

\

\begin{figure}[!h]
\begin{center}
 \scalebox{1.8}{\includegraphics{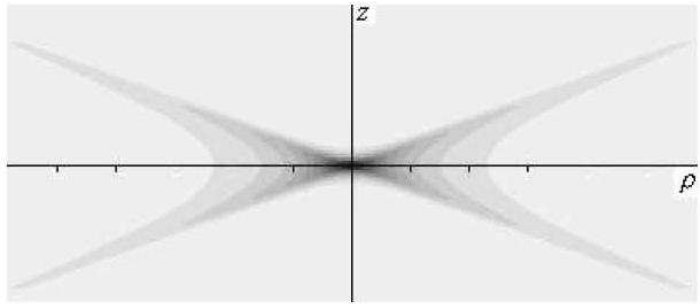}}  
\end{center}
\caption{The truncated X-waves considered in this paper, as
predicted by SR (all wave-equations have an intrinsic {\em
relativistic} structure!), must have a leading cone in addition to
the rear cone; such a leading cone having a role for the peak
stability\cite{Lu1}: For example, when producing a finite conic
wave truncated both in space and in time, the theory of SR
suggested to have recourse, in the {\em simplest} case, to a
dynamic antenna emitting a radiation cylindrically symmetrical in
space and symmetric in time, for a better approximation to what
Courant and Hilbert\cite{Courant} called an ``undistorted
progressing wave". See the following, in the text). } \label{fig7}
\end{figure}

(ii) Such pulses, even if their group-velocity is Superluminal, do
not contradict standard physics, having been found in what
precedes on the basis of the wave equations ---in particular, of
Maxwell equations\cite{Ziolk,PhysicaA}--- only.  Indeed, as we
shall better see in the historical Appendix, their existence can
be understood within special relativity
itself\cite{BarutMR,SupCh,PhysicaA,birdseye,JSTQE,RFG}, on the
basis of its ordinary Postulates\cite{Review}. Actually, let us
repeat it, they are fed by waves originating at the aperture
and carrying energy with the standard speed $c$ of the medium (the
light-velocity in the electromagnetic case, and the sound-velocity
in the acoustic\cite{Lu2} case).
 \ We can become convinced about the possibility of realizing Superluminal
X-shaped pulses by imagining the simple ideal case of a negligibly
sized Superluminal source $S$ endowed with speed $V>c$ in vacuum,
and emitting electromagnetic waves $W$ (each one traveling with
the invariant speed $c$). The electromagnetic waves will result to
be internally tangent to an enveloping cone $C$ having $S$ as its
vertex, and as its axis the propagation line $z$ of the
source\cite{Review,SupCh}: {\em This is completely analogous to
what happens for an airplane that moves in air with constant
supersonic speed}. \ The waves $W$ interfere mainly negatively
inside the cone $C$, and constructively on its surface. \ We can
place a plane detector orthogonally to $z$, and record magnitude
and direction of the $W$ waves that hit on it, as (cylindrically
symmetric) functions of position and of time. \ It will be enough,
then, to replace the plane detector with a plane antenna which
{\em emits} ---instead of recording--- exactly the same (axially
symmetric) space-time pattern of waves $W$, for constructing a
cone-shaped electromagnetic wave $C$ that will propagate with the
Superluminal speed $V$ (of course, without a source any longer at
its vertex): \ even if each wave $W$ travels with the invariant
speed $c$. \ {\em Once more, this is exactly what would happen in
the case of a supersonic airplane} (in which case $c$ is the sound
speed in air: for simplicity, assume the observer to be at rest
with respect to the air). \ For further details, see the quoted
references. \ Actually, by suitable superpositions, and {\em
interference}, of speed-$c$ waves, one can obtain pulses more and
more localized in the vertex region\cite{MRH}: That is, very
localized field-``blobs" traveling with Superluminal
group-velocity. \ This has nothing to do with the illusory
``scissors effect'', since such blobs, along their field-depth,
are a priori able, e.g., to get two successive (weak) detectors,
located at distance $L$, clicking after a time {\em smaller} than
$L/c$. Incidentally, an analysis of the above-mentioned case (that
of a supersonic plane or a Superluminal charge) led, as
expected\cite{Review}, to the simplest type of ``X-shaped
pulse"\cite{SupCh}. \ It might be useful, finally, to recall that
SR (even the wave-equations have an internal {\em relativistic}
structure!) implies considering also the forward cone: cf.
Fig.\ref{fig7}. The truncated X-waves considered in this paper,
for instance, must have a leading cone in addition to the rear
cone; such a leading cone having a role for the peak
stability\cite{Lu1}: For example, in the approximate case in which
we produce a finite conic wave truncated both in space and in
time, the theory of SR suggested the bi-conic shape (symmetrical
in space with respect to the vertex $S$) to be a better
approximation to a rigidly traveling wave (so that SR suggests to
have recourse to a dynamic antenna emitting a radiation
cylindrically symmetrical in space and symmetric in time, for a
better approximation to an ``undistorted progressing wave").

\

(iii)  Any solutions that depend on $z$ and on $t$ only through
the quantity $z-Vt$, like eq.(\ref{px}), will appear the same to
an observer traveling along $z$ with the speed $V$, whatever it be
(subluminal, luminal or Superluminal) the value of $V$. That is,
such a solution will propagate rigidly with speed $V$ (and in fact
there exist Superluminal, luminal and subluminal localized waves).
This further explains why our X-shaped pulses, after having been
produced, will travel almost rigidly at speed $V$ (in this case, a
faster-than-light group-velocity), all along their depth of field.
To be even clearer, let us consider a generic function, depending
on $z-Vt$ with $V>c$, and show, {\em by explicit calculations
involving the Maxwell equations only}, that it obeys the scalar
wave equation. \ Following Franco Selleri\cite{Selleri}, let us
consider, e.g., the wave function

\

\bb \Phi(x,y,z,t) \ug {\dis{\frac{a}
{{\sqrt{[b-ic(z-Vt)]^2+(V^2-c^2)(x^2+y^2)}}}}} \label{sellerieq10}
\ee

\

with $a$ and $b$ non-zero constants, $c$ the ordinary speed of
light, and $V>c$ \ [incidentally, this wave function is nothing
but the classic X-shaped wave in cartesian co-ordinates].  Let us
naively verify that it is a solution to the wave equation

\

\bb  \nabf^2 \Phi(x,y,z,t) - {\frac{1}{c^2}} \; {\frac
{\pa^2\Phi(x,y,z,t)}{\pa^2t}} \ug 0 \ . \label{sellerieq11} \ee

\

On putting

\bb R \, \equiv \,
\dis{{{\sqrt{[b-ic(z-Vt)]^2+(V^2-c^2)(x^2+y^2)}}}} \; ,
\label{sellerieq12} \ee

one can write $\Phi = a/R$ and evaluate the second derivatives

\

$$\frac{1}{a} \, \frac{\pa^2\Phi}{\pa^2z} \ug
\frac{c^2}{R^3} - \frac{3c^2}{R^5}\,[b-ic(z-Vt)]^2 \ ;$$

$$\frac{1}{a} \, \frac{\pa^2\Phi}{\pa^2x} \ug  - \frac{V^2-c^2}{R^3} \, + \,
3\left(V^2-c^2 \right)^2 \; \frac{x^2}{R^5} \ ;$$

$$\frac{1}{a} \, \frac{\pa^2\Phi}{\pa^2y} \ug - \frac{V^2-c^2}{R^3} \, + \,
3\left(V^2-c^2 \right)^2 \; \frac{y^2}{R^5} \;$$

$$\frac{1}{a} \, \frac{\pa^2\Phi}{\pa^2t} \ug
\frac{c^2V^2}{R^3} - \frac{3c^2V^2}{R^5}\,[b-ic(z-Vt)]^2 \ ;$$

\

wherefrom

\

$$\frac{1}{a} \, \left[ \frac{\pa^2\Phi}{\pa^2z} - \frac{1}{c^2}
\frac{\pa^2\Phi}{\pa^2t} \right] \ug - \frac{V^2-c^2}{R^3} +
3\left(V^2-c^2 \right)^2 \; \frac{[b-ic(z-Vt)]^2}{R^5} \ ,$$

and

$$\frac{1}{a} \, \left[ \frac{\pa^2\Phi}{\pa^2x} + \frac{\pa^2\Phi}{\pa^2y}
\right] \ug -2 \, \frac{V^2-c^2}{R^3} + 3\left(V^2-c^2 \right)^2
\; \frac{x^2+y^2}{R^5} \ .$$

\

From the last two equations, remembering the previous definition,
one finally gets

$$\frac{1}{a} \, \left[ {\frac{\pa^2\Phi}{\pa^2z}} + {\frac{\pa^2\Phi}{\pa^2x}}
+ {\frac{\pa^2\Phi}{\pa^2y}} - {\frac{1}{c^2}} \;
{\frac{\pa^2\Phi}{\pa^2t}} \right] \ug 0$$

that is nothing but the (d'Alembert) wave equation
(\ref{sellerieq11}), {\bf q.e.d.} \  In conclusion, function
$\Phi$ is a solution of the wave equation even if it does
obviously represent a pulse (Selleri says ``a signal") propagating
with Superluminal speed.

\

\h After the previous three important comments, let us go back to
our evaluations with regard to the X-type solutions to the wave
equations. \ Let us now consider in eq.(\ref{px}), for instance,
the particular frequency spectrum $F(\om)$ given by

\

\bb F(\om) \ug H(\om)\,\frac{a}{V}\,\,\,{\rm
exp}\left(-\frac{a}{V}\,\om\right) \; , \label{fx} \ee

\

where $H(\om)$ is the Heaviside step-function and $a$ a positive
constant. Then, eq.(\ref{px}) yields

\

\bb \psi(\rho,\zeta) \, \equiv \, X \ug \frac{a}{\sqrt{(a -
i\zeta)^2 + \left(\frac{V^2}{c^2}-1\right)\rho^2}} \;, \label{ox}
\ee

\

with $\zeta = z - Vt$.  This solution (\ref{ox}) is the well-known
ordinary, or ``classic", X-wave, which constitutes a simple
example of X-shaped pulse.\cite{Lu1,PhysicaA} \ Notice that
function (\ref{fx}) contains mainly low frequencies, so that the
classic X-wave is suitable for low frequencies only.

\begin{figure}[!h]
\begin{center}
 \scalebox{3.2}{\includegraphics{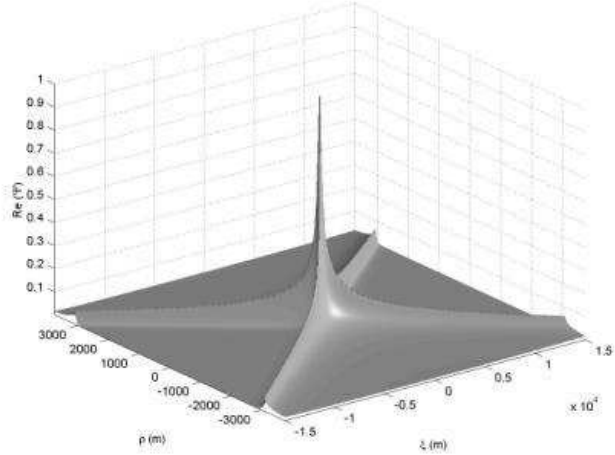}}  
\end{center}
\caption{Plot of the real part of the ordinary X-wave, evaluated
for $V=1.1\,c$ with $a=3\;$m .} \label{fig8}
\end{figure}

\h Figure \ref{fig8} does depict (the real part of) an ordinary
X-wave with $V=1.1\,c$ and $a=3\;$m.

\h Solutions (\ref{px}), and in particular the pulse (\ref{ox}),
have got an infinite field-depth, and an infinite energy as well.
Therefore, as it was done in the Bessel beam case, one should
proceed to truncated pulses, originating from a finite aperture.
Afterwards, our truncated pulses will keep their spatial shape
(and their speed) all along the depth of field

\bb Z \ug \frac{R}{\tan\theta} \; , \ee

where, as before, $R$ is the aperture radius and $\theta$ the
axicon angle.

\

{\em Some Further Observations} --- Let us put forth some further
observations.

\h It is not strictly correct to call non-diffractive the
localized waves, since diffraction affects, more or less, all
waves obeying eq.(\ref{eo1}). {\em However}, all localized waves
(both beams and pulses) possess the remarkable
``self-reconstruction" property: That is to say, the localized
waves, when diffracting during propagation, do immediately
re-build their shape\cite{Bouchal,Grunwald,Michel} (even after
obstacles with size much larger than the characteristic wave-lengths,
provided it is smaller ---as we know--- than the aperture size),
due to their particular spectral structure (as it will be shown
more in detail in other Chapters of the mentioned book [{\em Localized
Waves} (J.Wiley; in press]). In particular, the
``ideal localized waves'' (with infinite energy and depth of
field) are able to re-build themselves for an infinite time;
while, as we have seen, the finite-energy (truncated) ones can do
it, and thus resist the diffraction effects, only along a certain
field-depth...

\h Let us stress again that the interest of the localized waves
(especially from the point of view of applications) lies in the
circumstance that they are almost non-diffractive, rather than in
their group-velocity: From this point of view, Superluminal,
luminal, and subluminal localized solutions are equally
interesting and suited to important applications.

\begin{figure}[!h]    
\begin{center}
 \scalebox{2.1}{\includegraphics{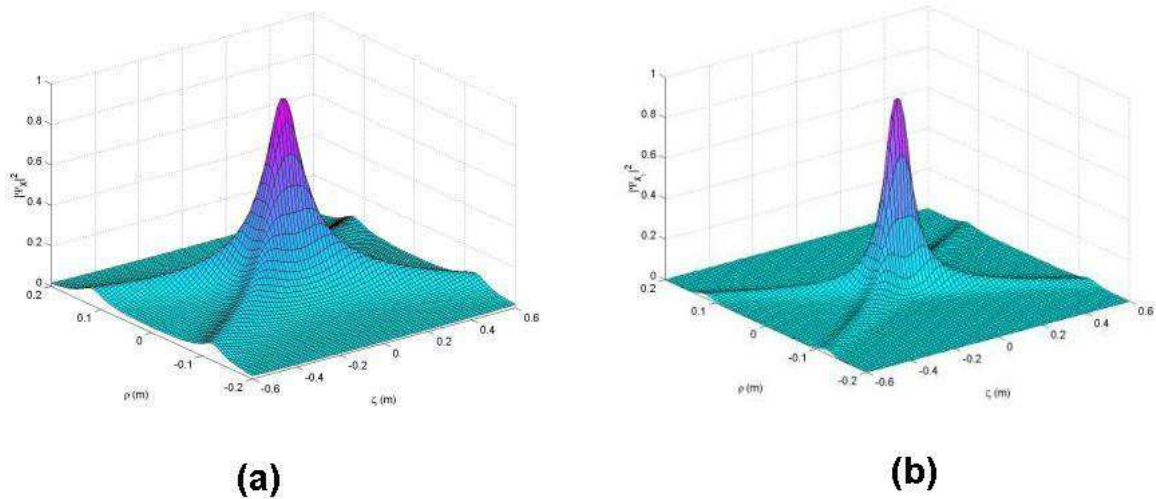}}    
 \end{center}
\caption{In Fig.(a) it is represented (in arbitrary units) the
square magnitude of the ``classic", $X$-shaped Superluminal
localized solution (SLS) to the wave equation, with $V=5c$ and
$a=0.1 \;$m. \ Families of infinite SLSs however exists, which
generalize the classic $X$-shaped solution; for instance, a family
of SLSs obtained\cite{MRH} by suitably differentiating the classic
X-wave: Fig.(b) depicts the first of them (corresponding to the
first differentiation) with the same parameters. \ As we said, the
successsive solutions in such a family are more and more localized
around their vertex.  Quantity $\rho$ is the distance in meters
from the propagation axis $z$, while quantity $\ze$ is the
``$V$-cone" variable[ref.\cite{MRH}] (still in meters) $\ze \equiv
z-Vt$, with $V \geq c$. \ Since all these solutions depend on $z$
only via the variable $\ze$, they propagate ``rigidly", i.e., as
we know, without distortion (and are called ``localized", or
non-dispersive, for such a reason).  Here we are assuming
propagation in the vacuum (or in a homogeneous medium). }
\label{fig9}
\end{figure}

\h Actually, the localized waves are not restricted to the
(X-shaped, Superluminal) ones corresponding to the integral
solution (\ref{px}) to the wave equation; and, as we were already
saying, three classes of localized pulses exist: the Superluminal
(with speed $V > c$, the luminal ($V=c$), and the subluminal
($V<c$) ones; all of them with, or without, axial symmetry, and,
in any case, corresponding to a unified, single mathematical
background. \ This issue will be touched again in the present
book. \ Incidentally, we have elsewhere addressed topics as \ (i)
the construction of infinite families of generalizations of the
classic X-shaped wave [with energy more and more concentrated
around the vertex: cf., e.g., Figs.\ref{fig9}, taken from
ref.\cite{MRH}]; as \ (ii) the behavior of some finite
total-energy Superluminal localized solutions (SLS); \ (iii) the
way for building up new series of SLS's to the Maxwell equations
suitable for arbitrary frequencies and bandwidths, as well as \
(iv) questions related with the case of dispersive media: In
Chapter 2 of the abovementioned book [{\em Localized Waves} (J.Wiley; in
press] we shall come back to some
(few) of those points. \ Let us add that X-shaped waves have been
easily produced also in nonlinear media\cite{Conti}, as a further
Chapter of the same volume will show.

\h A more technical introduction to the subject of localized waves
(particularly w.r.t. the Superluminal X-shaped ones) can be found
for instance in ref.\cite{JSTQE}.

\

\

\

\

\newpage

\

\

\

\centerline{{\bf  APPENDIX:}}

\centerline{A HISTORICAL (THEORETICAL AND EXPERIMENTAL) APPENDIX}

\centerline{====================================================}

\h In this mainly ``historical" Appendix, written as far as
possible in a (partially) self-consistent form, we shall first
refer ourselves, from the theoretical point of view, to the most
intriguing localized solutions to the wave equation: the
Superluminal ones (SLS), and in particular the X-shaped pulses. As
a start, we shall recall their geometrical interpretation
within SR. Afterwards, to help resolving possible doubts, we shall
seize the opportunity, given to us by this Appendix, for
presenting a bird's-eye view of the various {\em experimental}
sectors of physics in which Superluminal motions seem to appear:
In particular, of the experiments with evanescent waves (and/or
tunneling photons), and with the SLS's we are more interested in
here. In some parts of this Appendix the propagation-line is called
$x$, and no longer $z$, without originating, however, any
interpretation problems.

\

\section {INTRODUCTION OF THE APPENDIX}

The question of Superluminal ($V^{2}>c^{2}$) objects or waves has
a long story. Still in pre-relativistic times, one meets various
relevant papers, from those by J.J.Thomson to the interesting ones
by A.Sommerfeld. It is well-known, however, that with SR the
conviction spread out that the speed $c$ of light in vacuum was
the upper limit of any possible speed. For instance, R.C.Tolman in
1917 believed to have shown by his ``paradox'' that the existence
of particles endowed with speeds larger than $c$ would have
allowed sending information into the past. \ Our problem started
to be tackled again only in the fifties and sixties, in particular
after the papers\cite{ECGS1} by E.C.George Sudarshan et al., and,
later on\cite{Rev1974,North-Holl}, by one of the present authors
with R.Mignani et al., as well as  ---to confine ourselves at
present to the theoretical researches--- by H.C.Corben and others.
The first experimental attempts were performed by T.Alv\"{a}ger et
al.

\h We wish to face the still unusual issue of the possible
existence of Superluminal wavelets, and objects ---within standard
physics and SR, as we said--- since at least four different
experimental sectors of physics  {\em seem} to support such a
possibility [apparently confirming some long-standing theoretical
predictions\cite{Review,BarutMR,ECGS1,North-Holl}]. The
experimental review will be necessarily short, but we shall
provide the reader with further, enough bibliographical
information, limited for brevity's sake to the last century only
(i.e., up-dated till the year 2000 only). \

\

\section{APPENDIX: HISTORICAL RECOLLECTIONS - THEORY}

\begin{figure}[!h]
\begin{center}
 \scalebox{0.8}{\includegraphics{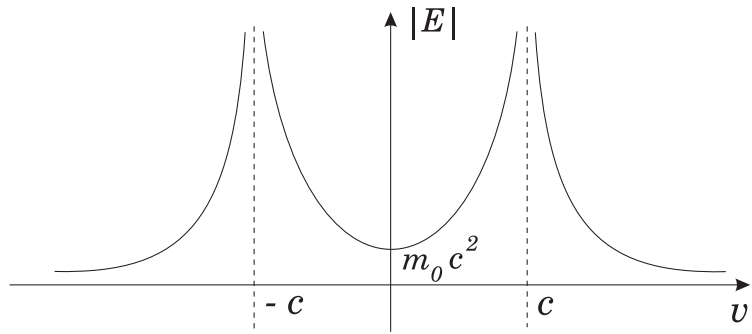}}    
\end{center}
\caption{Energy of a free object as a function of its
speed.\cite{ECGS1,
 Rev1974,Review}}
\label{fig10}
\end{figure}

\h A simple theoretical framework was long ago
proposed\cite{ECGS1,Review,Rev1974}, merely based on the
space-time geometrical methods of SR, which appears to incorporate
Superluminal waves and objects, and predict\cite{BarutMR} among
the others the Superluminal X-shaped waves, without violating the
Relativity principles. A suitable choice of the Postulates of SR
(equivalent of course to the other, more common, choices) is the
following one: (i) the standard Principle of Relativity; and (ii)
space-time homogeneity and space isotropy. \ It follows that one
and only one {\em invariant}  speed exists; and experience shows
that invariant speed to be the light-speed, $c$, in vacuum: The
essential role of $c$ in SR being just due to its invariance, and
not to the fact that it be a maximal, or minimal, speed. No sub-
or Super-luminal objects or pulses can be endowed with an
invariant speed: so that their speed cannot play in SR the same
essential role played the light-speed $c$ in vacuum. Indeed, the
speed $c$ turns out to be a {\em limiting} speed: but any limit
possesses two sides, and can be approached a priori both from
below and from above: See Fig.\ref{fig10}. \ As E.C.G.Sudarshan
put it, from the fact that no one could climb over the Himalayas
ranges, people of India cannot conclude that there are no people
North of the Himalayas; actually, speed-$c$  photons exist, which
are born, live and die just ``at the top of the mountain," without
any need for performing the impossible task of accelerating from
rest to the light-speed. \ [Actually, the ordinary formulation of
SR is restricted too much: For instance, even leaving Superluminal
speeds aside, it can be easily so widened as to include
antimatter\cite{Review,FoP87,RFG}].

\h An immediate consequence is that the quadratic form \ $c^{2}
\drm t^{2} - \drm \xbf^{2} \equiv \drm x_\mu \drm x^\mu$, called
$\drm s^{2}$, \ with $\drm \xbf^{2} \equiv \drm x^{2}+\drm
y^{2}+\drm z^{2}$, \ results to be invariant, {\em except for its
sign}. \ Quantity $\drm s^{2}$, let us emphasize, is the
four-dimensional length-element square, along the space-time path
of any object. \ In correspondence with the positive (negative)
sign, one gets the subluminal (Superluminal) Lorentz
``transformations" [LT]. The ordinary subluminal LTs are known to
leave, e.g., the quadratic forms $\drm x_\mu \drm x^\mu$, $\drm
p_\mu \drm p^\mu$ and  $\drm x_\mu \drm p^\mu$ exactly invariant,
where the $p_\mu$ are the component of the energy-impulse
four-vector; while the Superluminal LTs, by contrast, have to
change (only) the sign of such quadratic forms. This is enough for
deducing some important consequences, like the one that a
Superluminal charge has to behave as a magnetic monopole, in the
sense specified in ref.\cite{Review} and refs. therein.

\begin{figure}[!h]
\begin{center}
 \scalebox{1.76}{\includegraphics{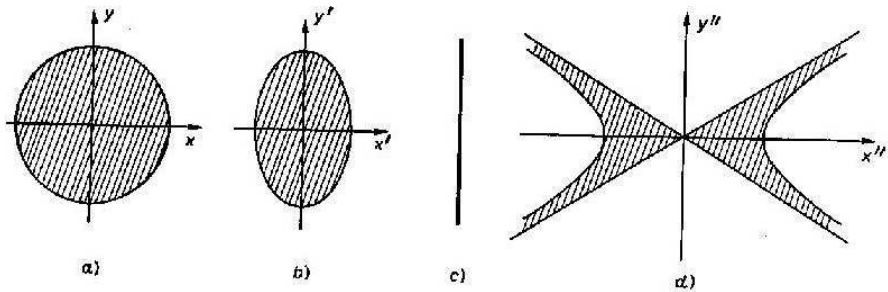}}  
\end{center}
\caption{An intrinsically spherical (or pointlike, at the limit)
object appears in the vacuum as an ellipsoid contracted along the
motion direction when endowed with a speed $v<c$. \ By contrast,
if endowed with a speed $V>c$ (even if the $c$-speed barrier
cannot be crossed, neither from the left nor from the right), it
would appear\cite{BarutMR,Review} no longer as a particle, but
rather as an ``X-shaped" wave travelling rigidly: Namely, as
occupying the region delimited by a double cone and a two-sheeted
hyperboloid ---or as a double cone, at the limit--, and moving
without distortion in the vacuum, or in a homogeneous medium, with
Superluminal speed $V$ [the cotangent square of the cone
semi-angle being $(V/c)^2-1$]. For simplicity, a space axis is
skipped. \ This figure is taken from refs.\cite{BarutMR,Review} }
\label{fig11}
\end{figure}

\h A more important consequence, for us, is ---see
Fig.\ref{fig11}--- that the simplest subluminal object, a
spherical particle at rest (which appears as ellipsoidal, due to
Lorentz contraction, at subluminal speeds $v$), will
appear\cite{BarutMR,Review,PhysicaA} as occupying the
cylindrically symmetrical region bounded by a two-sheeted rotation
hyperboloid and an indefinite double cone, as in Fig.11(d), for
Superluminal speeds $V$. In Fig.11 the motion is along the
$x$-axis. In the limiting case of a point-like particle, one
obtains only a double cone.

\begin{figure}[!h]
\begin{center}
 \scalebox{.96}{\includegraphics{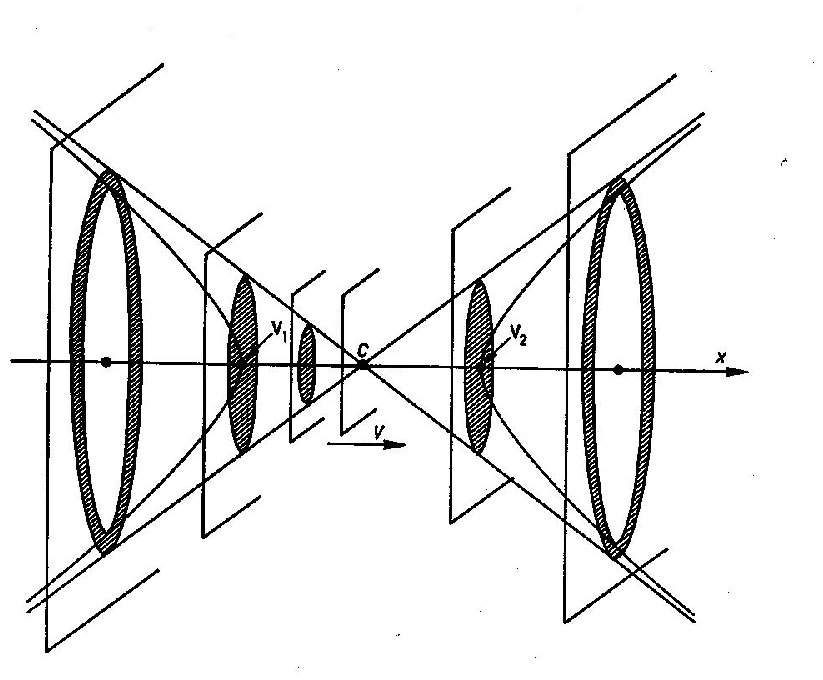}}  
\end{center}
\caption{Here we show the {\em intersections} of the Superluminal
object $T$ represented in Fig.11(d) with planes $P$ orthogonal to
its motion line (the $x$-axis). \ For simplicity, we assumed again
the object to be spherical in its rest-frame, and the cone vertex
$C$ to coincide with the origin $O$ for $t=0$. \ Such
intersections evolve in time so that the same pattern appears on a
second plane ---shifted by $\Delta x$--- after the time $\Delta t
= \Delta x / V$. \ On each plane, as time elapses, the
intersection is therefore predicted by (extended) SR to be a
circular ring which, for negative times, goes on shrinking until
it reduces to a circle and then to a point (for $t=0$); \
afterwards, such a point becomes again a circle and then a
circular ring that goes on
broadening\cite{BarutNM,Review,PhysicaA}. \ This picture is taken
from refs.\cite{BarutNM,Review}. \ [Notice that, if the object is
not spherical when at rest (but, e.g., is ellipsoidal in its own
rest-frame), then the axis of $T$  will no longer coincide with
$x$, but its direction will depend on the speed $V$ of the tachyon
itself]. \ For the case in which the space extension of the
Superluminal object $T$ is finite, see refs.\cite{BarutMR} }
\label{fig12}
\end{figure}

Such result is simply got by writing down the equation of the
{\em world-tube} of a subluminal particle, and transforming it by merely
changing sign to the quadratic forms entering that equation. Thus,
in 1980-1982, it was predicted\cite{BarutMR} that the simplest
Superluminal object appears (not as a particle, but as a field or
rather) as a wave: namely, as an ``X-shaped pulse", the cone
semi-angle $\al$ being given (with $c=1$) by \ ${\rm cotg} \al =
\sqrt{V^2 - 1}$. Such X-shaped pulses will move {\em rigidly} with
speed $V$ along their motion direction: In fact, any ``X-pulse"
can be regarded at each instant of time as the (Superluminal)
Lorentz transform of a spherical object, which of course moves in
vacuum ---or in a homogeneous medium--- without any deformation as
time elapses. \ The three-dimensional picture of Fig.11(d) appears
in Fig.\ref{fig12}, where its annular intersections with a
transverse plane are shown (cf. refs.\cite{BarutMR}). \ The
X-shaped waves here considered are merely the simplest ones: if
one starts not from an intrinsically spherical or point-like
object, but from a non-spherically symmetric particle, or from a
pulsating (contracting and dilating) sphere, or from a particle
oscillating back and forth along the motion direction, then their
Superluminal Lorentz transforms would result to be more and more
complicated. The above-seen ``X-waves", however, are typical for a
Superluminal object, so as the spherical or point-like shape is
typical, let us repeat, for a subluminal object.

\h Incidentally, it has been believed for a long time that
Superluminal objects would have allowed sending information into
the past; but such problems with causality seem to be solvable
within SR. Once SR is generalized in order to include Superluminal
objects or pulses, no signal traveling backwards in time is
apparently left.  For a solution of those causal paradoxes, see
refs.\cite{FoP87,RGF,ECGS1} and references therein.

\h For addressing the problem, even within this elementary
context, of the production of an X-shaped pulse like the one
depicted in Fig.\ref{fig12} (maybe truncated, in space and in
time, by use of a finite antenna radiating for a finite time), all
the considerations expounded under point (ii) of the subsection
{\em The Ordinary X-shaped Pulse} become in order: And, here, we
simply refer to them.  Those considerations, together with the
present ones (related, e.g., to Fig.\ref{fig12}), suggest the
simplest antenna to consist in a series of concentric annular
slits, or transducers [like in Fig.\ref{fig2}], which suitably
radiate following specific time patterns: See, e.g.,
refs.\cite{Mjosaatoappear} and refs. therein. \ Incidentally, the
above procedure can lead to a very simple type of X-shaped wave.

\ From the present point of view, it is rather interesting to note
that, during the last fifteen years, X-shaped waves have been {\em
actually} found as solutions to the Maxwell and to the wave
equations [let us recall that the form of any wave equations is
intrinsically relativistic]. {\em In order to see more deeply the
connection existing between what predicted by SR} (see, e.g.,
Figs.\ref{fig11},\ref{fig12}) {\em and the localized X-waves
mathematically, and experimentally, constructed in recent times,}
let us tackle below, in detail, the problem of the (X-shaped)
field created by a Superluminal electric charge\cite{SupCh}, by
following a paper recently appeared in Physical Review E.

\

\subsection{The particular X-shaped field associated with a Superluminal
charge}

It is well-known by now that Maxwell equations admit of
wavelet-type solutions endowed with arbitrary group-velocities ($0
< v_\grm < \infi$). \ We shall again confine ourselves, as above,
to the localized solutions, rigidly moving: and, more in
particular, to the Superluminal ones (SLS), the most interesting
of which resulted to be, as we have seen, X-shaped. \ The SLSs
have been actually produced in a number of experiments, always by
suitable interference of ordinary-speed waves. \ In this
subsection we show, by contrast, that even a Superluminal charge
creates an electromagnetic X-shaped wave, in agreement with what
predicted within SR\cite{BarutMR,Review}. Namely, on the basis of
Maxwell equations, one is able to evaluate the field associated
with a Superluminal charge (at least, under the rough
approximation of pointlikeness): as announced in what precedes, it
results to constitute a very simple example of {\em true} X-wave.

\h Indeed, the theory of SR, when based on the {\em ordinary}
Postulates but not restricted to subluminal waves and objects,
i.e., in its extended version, predicted the simplest X-shaped
wave to be the one corresponding to the electromagnetic field
created by a Superluminal charge\cite{Folman,SupCh}. \ It seems
really important evaluating such a field, at least approximately,
by following ref.\cite{SupCh}.

\

{\em The toy-model of a pointlike Superluminal charge} ---  Let us
start by considering, formally, a pointlike Superluminal charge,
even if the hypothesis of pointlikeness (already unacceptable in
the subluminal case) is totally inadequate in the Superluminal
case\cite{Review}. \ Then, let us consider the ordinary
vector-potential $A^\mu$ and a current density $j^\mu \equiv
(0,0,j_z;j^\orm)$ flowing in the $z$-direction (notice that the
motion line is here the axis $z$). On assuming the fields to be
generated by the sources only, one has that $A^\mu \equiv
(0,0,A_z;\phi)$, which, when adopting the Lorentz gauge, obeys the
equation $A^\mu = j^\mu$. \ We can write such non-homogeneous wave
equation in the cylindrical co-ordinates $(\rho,\theta,z;t)$; for
axial symmetry [which requires a priori that $A^\mu =
A^\mu(\rho,z;t)$], when choosing the ``$V$-cone variables" $\ze
\equ z-Vt; \ \eta \equ z+Vt \;$, with \ $V^2 > c^2$, we
arrive\cite{SupCh} to the equation

\

\bb \dis{\left[-\rho {\pa \over {\pa\rho}} \left(\rho {\pa \over
{\pa\rho}}\right) + \frac{1}{\ga^2} \frac{\pa^2}{\pa \ze^2} +
\frac{1}{\ga^2} \frac{\pa^2}{\pa \eta^2} - 4 \frac{\pa^2}{\pa\ze
\pa\eta} \right] \; A^\mu(\rho,\ze,\eta) \ug j^\mu(\rho,\ze,\eta)}
\ , \label{17} \ee

\

where $\mu$ assumes the two values $\mu = 3,0$ only, so that \
$A^\mu \equiv (0,0,A_z;\phi)$, \ and \ $\ga^2 \equiv
[V^2-1]^{-1}$. \ [Notice that, whenever convenient, we set $c=1$].
\ Let us now suppose $A^\mu$ to be actually independent of $\eta$,
namely, $A^\mu = A^\mu (\rho, \ze)$. \ Due to eq.(\ref{17}), we
shall have $j^\mu = j^\mu (\rho, \ze) \,$ too; and therefore $j_z
= V j^0$ (from the continuity equation), and $A_z = V \phi / c$
(from the Lorentz gauge). \ Then, by calling $\psi \equiv A_z$, we
end in two equations\cite{SupCh}, which allow us to analyze the
possibility and consequences of having a Superluminal pointlike
charge, $e$, traveling with constant speed $V$ along the $z$-axis
($\rho = 0$) in the positive direction, in which case $j_z = e\,
V\, \delta(\rho)/{\rho} \; \delta(\ze)$. \ Indeed, one of those
two equations becomes the hyperbolic equation

\

\bb \dis{\left[-\frac{1}{\rho}\frac{\pa}{\pa\rho}\left(\rho
\frac{\pa}{\pa\rho}\right) + \frac{1}{\ga^2}
\frac{\pa^2}{\pa\ze^2}\right] \, \psi \ug e V \, \frac{\delta(\rho
)}{\rho} \, \delta(\ze)} \label{18}\ee

\

which can be solved\cite{SupCh} in few steps. First, by applying
(with respect to the variable $\rho$) the Fourier-Bessel (FB)
transformation \ $f(x) \ug \dis{\int_0^{\infty} \Om f(\Om) J_0(\Om
x) \, \drm\Om}$, \ quantity $J_0(\Om x)$ being the ordinary
zero-order Bessel function. \ Second, by applying the ordinary
Fourier transformation with respect to the variable $\ze \,$
(going on, from $\ze$, to the variable $\om$). \ And, third, by
finally performing the corresponding {\em inverse} Fourier and FB
transformations. \ Afterwards, it is enough to have recourse to
formulae (3.723.9) and (6.671.7) of ref.\cite{[14]}, still with
$\ze \equiv z - Vt \,$, for being able to write down the solution
of eq.(\ref{18}) in the form

\

\hfill{$ \left\{\begin{array}{clr}

\psi(\rho,\ze) \ug 0 \ \ \ \ \ \ \ \ \ \ \ \ \ \ \ \ \ \ \ \ \ \ \
\ \ \ \ \ \ \
{\rm {for}} \ \ \ 0 < \ga\mid{\ze}\mid < \rho \\

\\

\dis{\psi(\rho,\ze) \ug \dis{e \frac{V}{\sqrt{\ze^2 -
\rho^2(V^2-1)}}}} \ \ \ \ \ {\rm {for}} \ \ \ 0 \le \rho <
\ga\mid{\ze}\mid \ .

\end{array} \right.
$\hfill} (25)

\

\begin{figure}[!h]
\begin{center}
 \scalebox{0.8}{\includegraphics{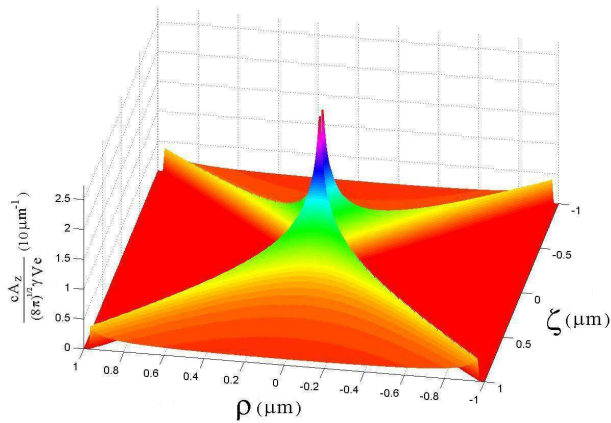}}  
\end{center}
\caption{Behaviour of the field $\psi \equiv A_z$  generated by a
charge supposed to be Superluminal, as a function of $\rho$ and
$\ze \equiv z-Vt$, evaluated for $\ga = 1$ (i.e., for $V = c
\sqrt{2}$): According to ref.\cite{SupCh} \ [Of course, we skipped
the points in which $\psi$ must diverge: namely, the vertex and
the cone surface]. }
\label{fig13}
\end{figure}

In Fig.\ref{fig13} we show our solution $A_z \equiv \psi$, as a
function of $\rho$ and $\ze$, evaluated for $\ga = 1$ (i.e., for
$V = c \sqrt{2}$). Of course, we skipped the points in which $A_z$
{\em must} diverge, namely the vertex and the cone surface.

\h For comparison, one may recall that the {\em classic} X-shaped
solution\cite{Lu1} of the {\em homogeneous} wave-equation ---which
is shown, e.g., in Figs.\ref{fig8},\ref{fig9},\ref{fig12}--- has
the form (with $a > 0$):

\

\hfill{$ \dis{X \ug {\frac{V}{\sqrt{(a-i\ze)^2 + \rho^2(V^2-1)}}}}
\; . $\hfill} (26)

\

The second one of eqs.(25) includes expression (26), given by the
spectral parameter\cite{MRH,meiodisp} $a = 0$, which indeed
corresponds to the non-homogeneous case [the fact that for $a=0$
these equations differ for an imaginary unit will be discussed
elsewhere].

\begin{figure}[!h]
\begin{center}
 \scalebox{1.08}{\includegraphics{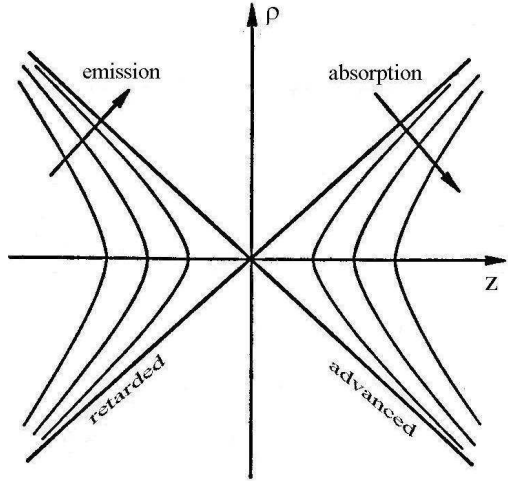}}  
\end{center}
\caption{The spherical equipotential surfaces of the electrostatic
field created by a charge at rest get transformed into two-sheeted
rotation-hyperboloids, contained inside an unlimited double-cone,
when the charge travels at Superluminal speed (cf.
refs.\cite{SupCh,Review}). This figures shows, among the others,
that a Superluminal charge traveling at constant speed, in a
homogeneous medium like the vacuum, does {\em not} lose
energy\cite{Folman}. \ Let us mention, incidentally, that this
double cone has nothing to do with the Cherenkov cone. [The
present picture is a reproduction of figure 27, appeared in 1986
at page 80 of ref.\cite{Review}]. }
\label{fig14}
\end{figure}

\h It is rather important, at this point, to notice that such a
solution, eq.(25), does represent a wave existing only inside the
(unlimited) double cone $\Ccal$ generated by the rotation around
the $z$-axis of the straight lines $\rho = \pm \ga\ze$: \ This too
is in full agreement with the predictions of the extended theory
of SR. \ For the explicit evaluation of the electromagnetic fields
generated by the Superluminal charge (and of their boundary values
and conditions) we confine ourselves here to merely quoting
ref.\cite{SupCh}. \ Incidentally, the same results found by
following the above procedure can be obtained by starting from the
four-potential associated with a subluminal charge (e.g., an
electric charge at rest), and then applying to it the suitable
Superluminal Lorentz ``transformation". \ One should also notice
that this double cone does not have much to do with the Cherenkov
cone\cite{Review,Folman}; and a Superluminal charge traveling at
constant speed, in the vacuum, does {\em not} lose energy: See,
e.g., Fig.\ref{fig14} [which reproduces figure 27 at page 80 of
ref.\cite{Review}].

\h Outside the cone $\Ccal$, i.e., for $0 < \ga\mid{\ze}\mid <
\rho$, we get as expected no field, so that one meets a field
discontinuity when crossing the double-cone surface. Nevertheless,
the boundary conditions imposed by Maxwell equations are satisfied
by our solution (25), since at each point of the cone surface the
electric and the magnetic field are both tangent to the cone: also
for a discussion of this point we refer to quotation\cite{SupCh}.

\h Here, let us stress that, when $V \ra \infi; \ \ga \ra 0$, the
electric field tends to vanish, while the magnetic field tends to
the value $H_\phi = -\pi e / \rho^2$: This does agree once more
with what expected from extended SR, which predicts Superluminal
charges to behave (in a sense) as magnetic monopoles. In the
present contribution we can only mention such a circumstance, and
refer to citations \cite{Rev1974,Review,North-Holl,chmonPLA}, and
papers quoted therein.

\

\

\section{APPENDIX: A GLANCE AT THE EXPERIMENTAL STATE-OF-THE-ART}

\h Extended relativity can allow a better understanding of many
aspects also of {\em ordinary} physics\cite{Review}, even if
Superluminal objects (tachyons) did not exist in our cosmos as
asymptotically free objects. \ Anyway, at least three or four
different experimental sectors of physics seem to suggest the
possible existence of faster-than-light motions, or, at least, of
Superluminal group-velocities. \ We are going to put forth in the
following some information about the experimental results obtained
in {\em two} of those different physics sectors, with a mere
mention of the others.

\

\h {\em Neutrinos} -- First: A long series of experiments, started
in 1971, seems to show that the square ${m_{0}}^{2}$ of the mass
$m_{0}$ of muon-neutrinos, and more recently of electron-neutrinos
too, is negative; which, if confirmed, would mean that (when using
a na\"{i}ve language, commonly adopted) such neutrinos possess an
``imaginary mass'' and are therefore tachyonic, or mainly
tachyonic.\cite{neutrinos,Review,Giannetto}  \ [In extended SR,
the dispersion relation for a free Superluminal object becomes \
$\om^2-\kbf^2=-\Om^2$, \ or \ $E^2-\imp^2=-m_\orm^{2}$, \ and
there is {\em no} need at all, therefore, of imaginary masses].

\

\h {\em Galactic Micro-quasars} -- Second: As to the {\em
apparent} Superluminal expansions observed in the core of
quasars\cite{quasars} and, recently, in the so-called galactic
micro-quasars\cite{microquasars}, we shall not really deal with
that problem, too far from the other topics of this paper; without
mentioning that for those astronomical observations there exist
orthodox interpretations, based on ref.\cite{Rees}, that are still
accepted by the majority of the astrophysicists. \ For a
theoretical discussion, see ref.\cite{Rodono}.  Here, let us
mention only that simple geometrical considerations in Minkowski
space show that a {\em single} Superluminal source of light would
appear\cite{Rodono,Review}: \ (i) initially, in the ``optical
boom'' phase (analogous to the acoustic ``boom'' produced by an
airplane traveling with constant supersonic speed), as an intense
source which suddenly comes into view; and which, afterwards, \
(ii) seems to split into TWO objects receding one from the other
with speed \ $V>2c$ [all of this being similar to what is actually
observed, according to refs.\cite{microquasars}].

\

\begin{figure}[!h]
\begin{center}
 \scalebox{1.43}{\includegraphics{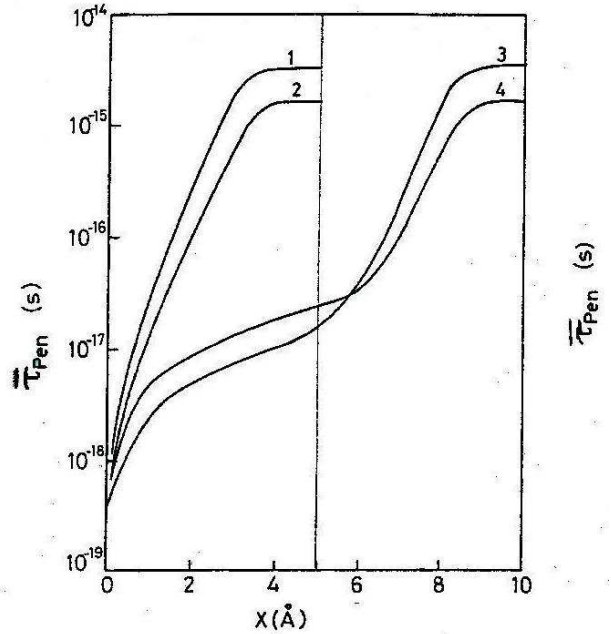}}  
\end{center}
\caption{Behaviour of the average ``penetration time" (in seconds)
spent by a tunnelling wavepacket, as a function of the penetration
depth (in {\aa}ngstroms) down a potential barrier (from Olkhovsky
et al., ref.\cite{Physique}). \ According to the predictions of
quantum mechanics, the wavepacket speed inside the barrier
increases in an unlimited way for opaque barriers; and the total
tunnelling time does {\em not} depend on the barrier
width.\cite{Reports1,Physique} } \label{fig15}
\end{figure}

\h {\em Evanescent waves and ``tunneling photons''} -- Third:
Within quantum mechanics (and precisely in the {\em tunneling}
processes), it had been shown that the tunneling time ---firstly
evaluated as a simple Wigner's ``phase time'' and later on
calculated through the analysis of the wavepacket behavior--- does
not depend\cite{Reports1} on the barrier width in the case of
opaque barriers (``Hartman effect''). This implies Superluminal
and arbitrarily large group-velocities $V$ inside long enough
barriers: see Fig.\ref{fig15}. \

\begin{figure}[!h]
\begin{center}
 \scalebox{1.6}{\includegraphics{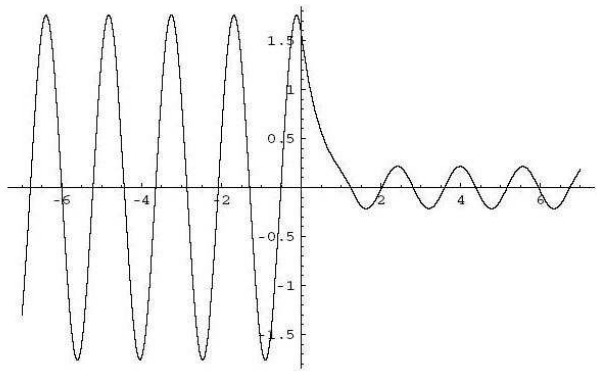}}  
\end{center}
\caption{Picture of the damping taking place inside a barrier
(from ref.\cite{RFG}): this damping does reduce the amplitude of
the tunnelling wavepacket, imposing a practical limit on the
barrier length .}
\label{fig16}
\end{figure}

Experiments that may verify this prediction by, say, electrons or
neutrons are difficult and rare\cite{Reports2,ORZ2}. Luckily
enough, however, the Schroedinger equation in the presence of a
potential barrier is mathematically identical to the Helmholtz
equation for an electromagnetic wave propagating, for instance,
down a metallic waveguide (along the $x$-axis): as shown, e.g., in
refs.\cite{ref[13]}; \ and a barrier height $U$ bigger than the
electron energy $E$ corresponds (for a given wave frequency) to a
waveguide of transverse size lower than a cut-off value. A segment
of ``undersized" guide ---to go on with our example--- does
therefore behave as a barrier for the wave (photonic barrier), as
well as any other photonic band-gap filters. \ The wave assumes
therein ---like a particle inside a quantum barrier--- an
imaginary momentum or wavenumber and, as a consequence, results
exponentially damped along $x$ [see, e.g. Fig.\ref{fig16}]: It
becomes an {\em evanescent} wave (going back to normal
propagation, even if with reduced amplitude, when the narrowing
ends and the guide returns to its initial transverse size). \
Thus, a tunneling experiment can be simulated by having recourse
to evanescent waves (for which the concept of group velocity can
be properly extended: see the first one of refs.\cite{RFG}).

\begin{figure}[!h]
\begin{center}
 \scalebox{1.95}{\includegraphics{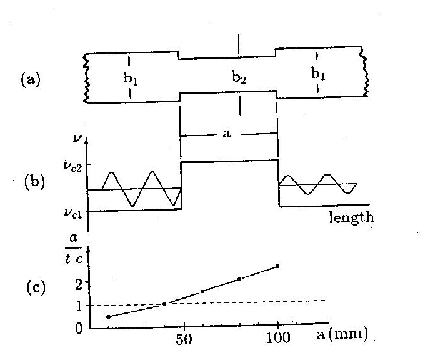}}  
\end{center}
\caption{Simulation of tunnelling by experiments with evanescent
{\em classical} waves (see the text), which were predicted to be
Superluminal also on the basis of extended SR\cite{ref19}. The
figure shows one of the measurement results by Nimtz et
al.\cite{Nimtz}; that is, the average beam speed while crossing
the evanescent region ( = segment of undersized waveguide, or
``barrier") as a function of its length. As theoretically
predicted\cite{Reports1,ref19}, such an average speed exceeds $c$
for long enough ``barriers". \ Further results appeared in
ref.\cite{Longhi}, and are reported below: see Figs.\ref{fig20}
and \ref{fig21} in the following .} \label{fig17}
\end{figure}

\begin{figure}[!h]
\begin{center}
 \scalebox{0.53}{\includegraphics{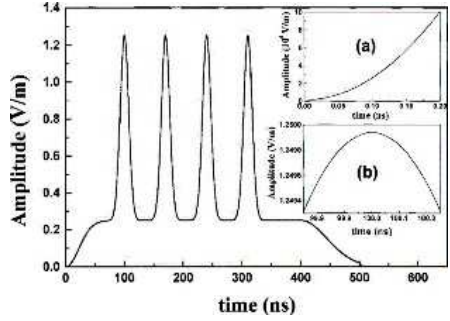}}  
\end{center}
\caption{The delay of a wavepacket crossing a barrier (cf., e.g.,
Fig.\ref{fig17} is due to the initial discontinuity. We then
performed suitable numerical simulations\cite{Barbero} by
considering an (indefinite) {\em undersized} waveguide, and
therefore eliminating any geometric discontinuity in its
cross-section.   \ This figure shows the envelope of the initial
signal. \ Inset (a) depicts in detail the initial part of this
signal as a function of time, while inset (b) depicts the gaussian
pulse peak centered at $t = 100$ ns .} \label{fig18}
\end{figure}

\begin{figure}[!h]
\begin{center}
 \scalebox{0.52}{\includegraphics{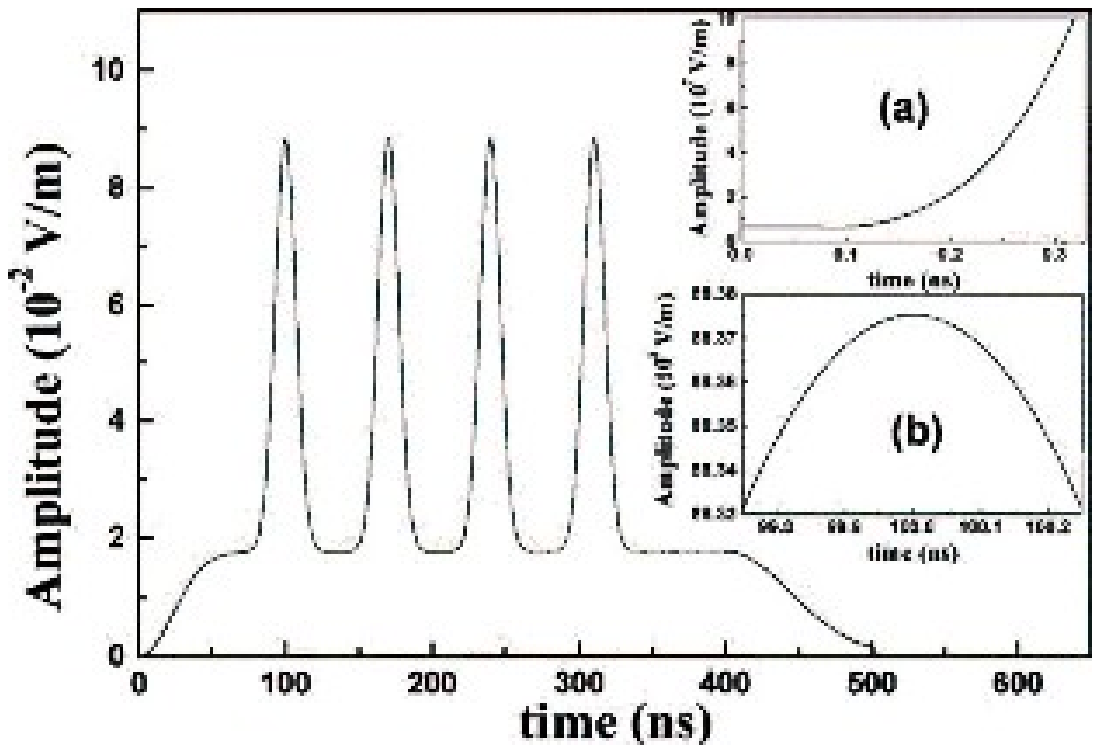}} 
\end{center}
\caption{Envelope of the signal in the previous figure
(Fig.\ref{fig18}) after having traveled a distance $L = 32.96$ mm
through the mentioned undersized waveguide. \ Inset (a) shows in
detail the initial part (in time) of such arriving signal, while
inset (b) shows the peak of the gaussian pulse that had been
initially modulated by centering it at $t = 100$ ns. \ One can see
that its propagation took {\em zero} time, so that the signal
traveled with infinite speed. \ The numerical simulation has been
based on Maxwell equations only. \ Going on from Fig.18 to Fig.19
one verifies that the signal strongly lowered its amplitute:
However, the width of each peak did not change (and this might
have some relevance when thinking of a Morse alphabet
``transmission": see the text) .} \label{fig19}
\end{figure}

\h The fact that evanescent waves travel with Superluminal speeds
(cf., e.g., Fig.\ref{fig17}) has been actually {\em verified} in a
series of famous experiments. Namely, various experiments,
performed since 1992 onwards by G.Nimtz et al. in
Cologne\cite{Nimtz}, by R.Chiao, P.G.Kwiat and A.Steinberg at
Berkeley\cite{Chiao}, by A.Ranfagni and colleagues in
Florence\cite{Ranfagni}, and by others in Vienna, Orsay, Rennes,
etcetera, verified that ``tunneling photons" travel with
Superluminal group velocities [Such experiments raised a great
deal of interest\cite{Chiao2}, also within the non-specialized
press, and were reported in Scientific American, Nature, New
Scientist, etc.]. \ Let us add that also extended SR had
predicted\cite{ref<} evanescent waves to be endowed with
faster-than-$c$ speeds; the whole matter appears to be therefore
theoretically selfconsistent. \ The debate in the current
literature does not refer to the experimental results (which can
be correctly reproduced even by numerical
simulations\cite{Barbero,Brodowsky} based on Maxwell equations
only: Cf. Figs.\ref{fig18},\ref{fig19}), but rather to the
question whether they allow, or do {\em not} allow, sending
signals or information with Superluminal speed (see, e.g.,
refs.\cite{debates}).

In the above-mentioned experiments one meets a substantial
attenuation of the considered pulses ---cf. Fig.\ref{fig16}---
during tunneling (or during propagation in an absorbing medium):
However, by employing ``gain doublets", it has been recently
reported the observation of undistorted pulses propagating with
Superluminal group-velocity with a {\em small} change in amplitude
(see, e.g., ref.\cite{Wang}).

\begin{figure}[!h]
\begin{center}
 \scalebox{0.8}{\includegraphics{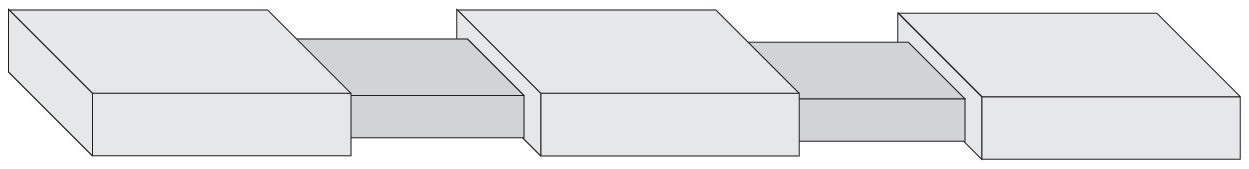}}  
\end{center}
\caption{Very interesting experiments have been performed with TWO
successive barriers, i.e., with two evanescence regions: For
example, with two gratings in an optical fiber. \ This
figure\cite{RFG} refers to the interesting experiment\cite{Nimtz3}
performed with microwaves traveling along a metallic waveguide:
the waveguide being endowed with {\em two} classical barriers
(undersized guide segments). See the text .} \label{fig20}
\end{figure}

\h  Let us emphasize that some of the most interesting experiments
of this series seem to be the ones with TWO or more ``barriers"
(e.g., with two gratings in an optical fiber, or with two segments
of undersized waveguide separated by a piece of normal-sized
waveguide: Fig.\ref{fig20}).

\

\begin{figure}[!h]
\begin{center}
 \scalebox{3.8}{\includegraphics{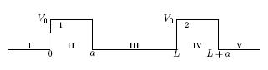}}  
\end{center}
\caption{Scheme of the (non-resonant) tunnelling process, through
{\em two} successive (opaque) quantum barriers. Far from
resonances, the (total) {\em phase time} for tunnelling through
the two potential barriers does depend neither on the barrier
widths {\em nor on the distance between the barriers}
(``generalized Hartman effect")\cite{twobar,Reports2,predic2b}. \
See the text.} \label{fig21}
\end{figure}

For suitable frequency bands ---namely, for ``tunneling" far from
resonances---, it was found by us that the total crossing time
does not depend on the length of the intermediate (normal) guide:
that is, that the beam speed along it is
infinite\cite{twobar,Nimtz3,Reports2}. \ This does agree with what
predicted by Quantum Mechanics for the non-resonant tunneling
through two successive opaque barriers\cite{twobar}: Fig.\ref{fig21}. \
Such a prediction has been verified first theoretically, by
Y.Aharonov et al.\cite{twobar}, and then, a second time,
experimentally: by taking advantage of the circumstance that
evanescence regions can consist in a variety of photonic band-gap
materials or gratings (from multilayer dielectric mirrors, or
semiconductors, to photonic crystals). Indeed, the best

\setcounter{figure}{21}
\begin{figure}[!h]
\begin{center}
 \scalebox{1.65}{\includegraphics{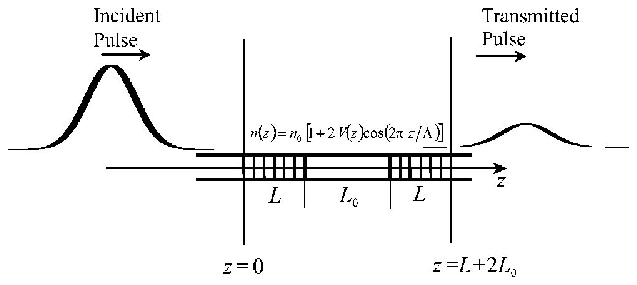}} 
\end{center}
\caption{Realization of the quantum-theoretical set-up represented
in Fig.\ref{fig21}, by using, as classical (photonic) barriers,
two gratings in an optical fiber\cite{predic2b}. \ The
corresponding experiment has been performed by Longhi et
al.\cite{Longhi} }
\label{fig22}             
\end{figure}

experimental confirmation has come by having recourse to two
gratings in an optical fiber\cite{Longhi}: see Figs.\ref{fig22}
and \ref{fig23}; in particular, the rather peculiar (and quite
interesting) results represented by the latter.

\begin{figure}[!h]
\begin{center}
 \scalebox{.5}{\includegraphics{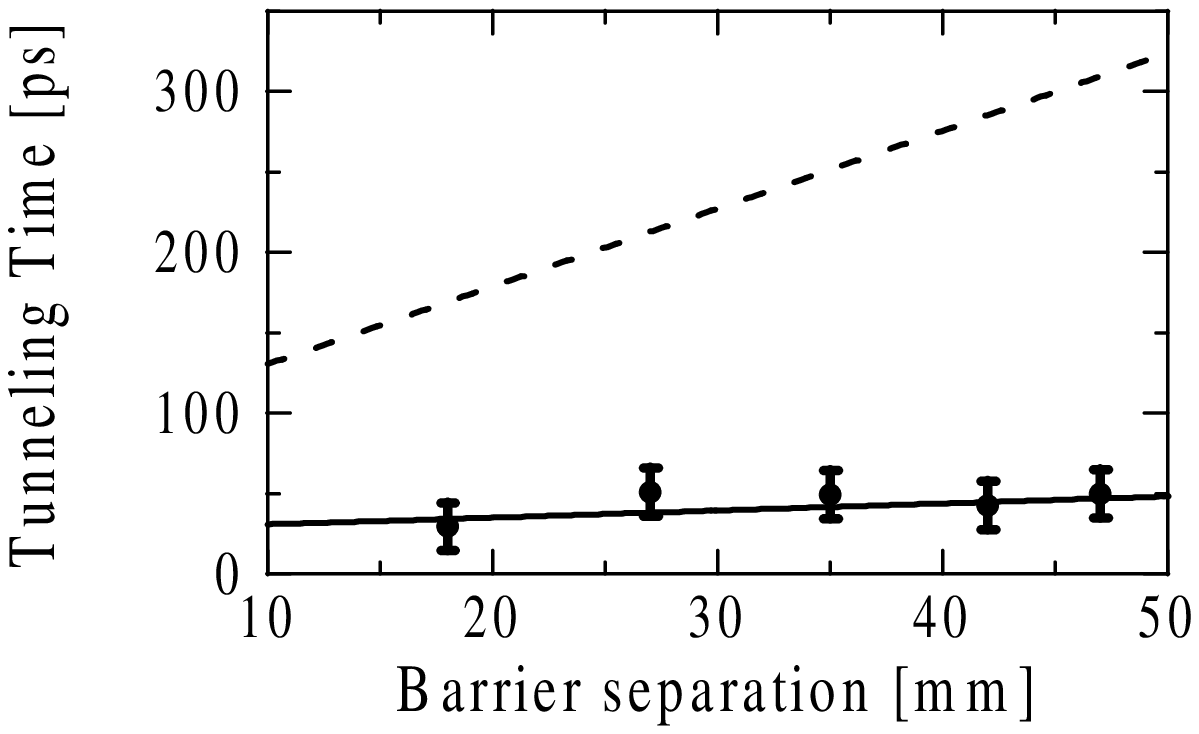}} 
\caption{Off-resonance tunnelling time versus barrier separation
for the rectangular symmetric DB FBG structure considered in
ref.\cite{Longhi} (see Fig.22). The solid line is the theoretical
prediction based on group delay calculations; {\em the dots are
the experimental points} as obtained by time delay measurements
[the dashed curve is the expected transit time from input to
output planes for a pulse tuned far away from the stopband of the
FBGs]. \ The experimental results\cite{Longhi} do confirm ---as
well as the early ones in refs.\cite{Nimtz3}]--- the theoretical
prediction  of a ``generalized Hartman Effect": in particular, the
independence of the total tunnelling time from the distance
between the two barriers. }
\end{center}
\label{fig23}
\end{figure}

\

\h  We cannot skip a further topic ---which, being delicate,
should not appear, probably, in a brief overview like this---
since it is presently arising more and more interest\cite{Wang}. \
Even if all the ordinary causal paradoxes seem to be
solvable\cite{FoP87,Review,RFG}, nevertheless one has to bear in
mind that (whenever it is met an object, ${\cal O}$, traveling
with Superluminal speed) one may have to deal with negative
contributions to the {\em tunneling
times\/}\cite{negativeTh,Review,Reports2}: and this should not be
regarded as unphysical. In fact, whenever an ``object'' (particle,
electromagnetic pulse,,...) ${\cal O}$ {\em
overcomes\/}\cite{FoP87,Review} the infinite speed with respect to
a certain observer, it will afterwards appear to the same observer
as the ``{\em anti}-object'' $\overline{{\cal O}} $ traveling in
the opposite {\em space} direction\cite{ECGS1,Review,FoP87}. \ For
instance, when going on from the lab to a frame ${\cal F}$ moving
in the {\em same} direction as the particles or waves entering the
barrier region, the object ${\cal O}$ penetrating through the
final part of the barrier (with almost infinite
speed\cite{Physique,Reports1,Barbero,Reports2}, like in Figs.15)
will appear in the frame ${\cal F}$ as an anti-object
$\overline{{\cal O}}$ crossing that portion of the barrier {\em in
the opposite space-direction\/}\cite{FoP87,Review,ECGS1}. In the
new frame ${\cal F}$, therefore, such anti-object $\overline{{\cal
O}}$ would yield a {\em negative} contribution to the tunneling
time: which could even result, in total, to be negative. \ For any
clarifications, see the quoted references. \ Let us stress, here,
that even the appearance of such negative times has been predicted
within SR itself, on the basis of its ordinary postulates; and
recently confirmed by quantum-theoretical evaluations
too\cite{Reports2,Petrillo}. \ (In the case of a non-polarized
beam,, the wave anti-packet coincides with the initial wave
packet; if a photon is however endowed with helicity $\lambda
=+1$, the anti-photon will bear the opposite helicity $\lambda
=-1$). \ From the theoretical point of view, besides the
above-quoted papers (in particular refs.\cite{Reports2,Reports1}),
see more specifically refs.\cite{negativeTh2}. \ On the (very
interesting!) {\em experimental} side, see the intriguing papers
\cite{negativeTh2}.

\h  Let us {\em add} here that, via quantum interference effects,
it is possible to obtain dielectrics with refraction indices very
rapidly varying as a function of frequency, also in three-level
atomic systems, with almost complete absence of light absorption
(i.e., with quantum induced transparency)\cite{Alzetta}. \ The
group velocity of a light pulse propagating in such a medium can
decrease to very low values, either positive or negative, with
{\em no} pulse distortion. \ It is known that experiments have
been performed both in atomic samples at room temperature, and in
Bose-Einstein condensates, which showed the possibility of
reducing the speed of light to a few meters per second. \ Similar,
but negative group velocities, implying a propagation with
Superluminal speeds thousands of time higher than the previously
mentioned ones, have been recently predicted also in the presence
of such an ``electromagnetically induced transparency'', for light
moving in a rubidium condensate\cite{Artoni}. \ Finally, let us
recall that faster-than-$c$ propagation of light pulses can be
(and has been, in same cases) observed also by taking advantage of the
anomalous dispersion near an absorbing line, or nonlinear and
linear gain lines ---as already seen---, or nondispersive
dielectric media, or inverted two-level media, as well as of some
parametric processes in nonlinear optics (cf., e.g., G.Kurizki et
al.'s works).

\

{\bf D)} \ {\bf Superluminal Localized Solutions (SLS) to the wave
equations. The ``X-shaped waves"} -- The fourth sector (to leave
aside the others) is not less important. It came into fashion
again, when it was rediscovered in a series of remarkable works
that any wave equation ---to fix the ideas, let us think of the
electromagnetic case--- admit also solutions as much sub-luminal
as Super-luminal (besides the luminal ones, having speed $c/n$). \
Let us recall, indeed, that, starting from pioneering works as
H.Bateman's, it had slowly become known that all wave equations
admit soliton-like (or rather wavelet-type) solutions with
sub-luminal group velocities. \ Subsequently, also Superluminal
solutions started to be written down (in one case\cite{Barut2}
just by the mere application of a Superluminal Lorentz
``transformation"\cite{Review}).

\begin{figure}[!h]
\begin{center}
 \scalebox{0.8}{\includegraphics{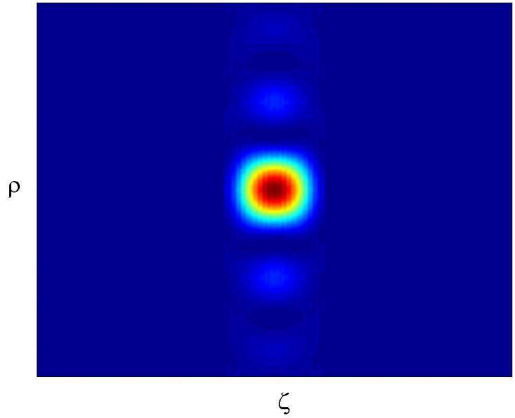}} 
\end{center}
\caption{The wave equations possess pulse-type solutions that, in
the subluminal case, are ball-like, in agreement with
Fig.\ref{fig11}.  \ For comments, see the text. }
\label{fig24}
\end{figure}

\h  As we know, a remarkable feature of some new solutions of
these (which attracted much attention for their possible
applications) is that they propagate as localized, non-dispersive
pulses, also because of their self-reconstruction property. \  It
is easy to realize the practical importance, for instance, of a
radio transmission carried out by localized beams, independently
of their speed; but non-dispersive wave packets can be of use even
in theoretical physics for a reasonable representation of
elementary particles; and so on. \ Incidentally, from the point of
view of elementary particles, it can be a source of meditation the
fact that the wave equations possess pulse-type solutions that, in
the subluminal case, are ball-like (cf. Fig.\ref{fig24}): this can
have a bearing on the corpuscle/wave duality problem met in
quantum physics (besides agreeing, e.g., with Fig.\ref{fig11}).

\h  At the cost of repeating ourselves, let us emphasize once more
that, within extended SR, since 1980 it had been found that
---whilst the simplest subluminal object conceivable is a small
sphere, or a point in the limiting case--- the simplest
Superluminal objects results by contrast to be (see
refs.\cite{BarutMR}, and Figs.\ref{fig11} and \ref{fig12} of this
paper) an ``X-shaped'' wave, or a double cone as its limit, which
moreover travels without deforming ---i.e., rigidly--- in a
homogeneous medium. \ It is not without meaning that the most
interesting localized solutions to the wave equations happened to
be just the Superluminal ones, and with a shape of that kind. \
Even more, since from Maxwell equations under simple hypotheses
one goes on to the usual {\em scalar} wave equation for each
electric or magnetic field component, one expected the same
solutions to exist also in the field of acoustic waves, of seismic
waves, and of gravitational waves too: and this has already been
demonstrated in the literature for the acoustic case. \ Actually,
such pulses (as suitable superpositions of Bessel beams) were
mathematically constructed for the first time, by Lu et al. {\em
in Acoustics\/}: and were then called ``X-waves'' or rather
X-shaped waves.

\begin{figure}[!h]
\begin{center}
 \scalebox{1.2}{\includegraphics{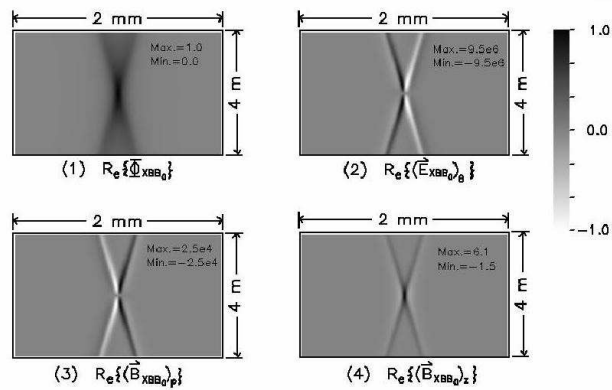}} 
\end{center}
\caption{Real part of the Hertz potential and of the field
components of the localized electromagnetic (``classic", axially
symmetric) X-shaped wave predicted, and first mathematically
constructed for the electromagnetic case, in refs.\cite{PhysicaA}.
\ For the meaning of the various panels, see the quoted
references. \ The dimension of each panel is 4 m (in the radial
direction) ${\times}$ 2 mm (in the propagation direction). \ [The
values shown on the right-top corner of each panel represent the
maxima and the minima of the images before normalization for
display (MKSA units)] .}
\label{fig25}
\end{figure}

\h  It is indeed important for us that the X-shaped waves have
been indeed produced in experiments, both with acoustic and with
electromagnetic waves; that is, X-pulses were produced that, in
their medium, travel undistorted with a speed larger than sound,
in the first case, and than light, in the second case. \ In
Acoustics, the first experiment was performed by Lu et al.
themselves in 1992, at the Mayo Clinic (and their papers received
the first 1992 IEEE award). \ In the electromagnetic case,
certainly more intriguing, Superluminal localized X-shaped
solutions were first mathematically constructed (cf., e.g.,
Fig.\ref{fig25}) in refs.\cite{PhysicaA}, and later on

\begin{figure}[!h]
\begin{center}
 \scalebox{1.82}{\includegraphics{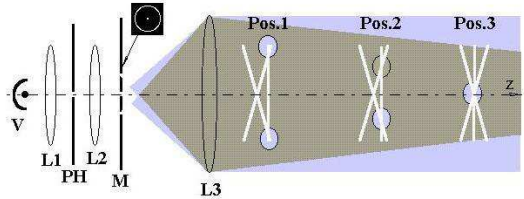}}  
\end{center}
\caption{Scheme of the experiment by Saari et al., who announced
(PRL of 24 Nov.1997) the production in optics of the beams
depicted in the previous Fig.\ref{fig25}. \  In the present figure
one can see what it was shown by the experimental results: Namely,
that the ``X-shaped" waves are Superluminal: indeed, they, running
after plane waves (the latter regularly traveling with speed $c$),
do catch up with the considered plane waves. \ An analogous
experiment has been later on performed with microwaves at Florence
by Mugnai, Ranfagni and Ruggeri (PRL of 22 May 2000). }
\label{fig26}
\end{figure}

\begin{figure}[!h]
\begin{center}
 \scalebox{2.6}{\includegraphics{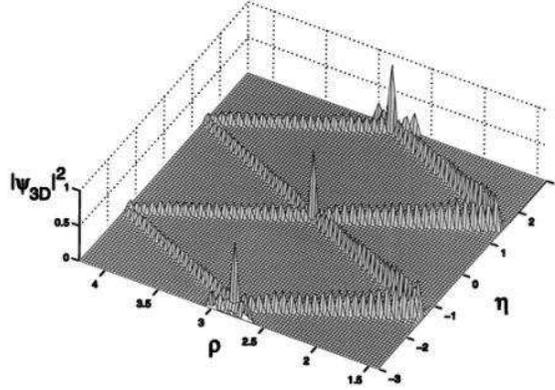}} 
\caption{In this figure a couple of elements are depicted of one
of the trains of X-shaped pulses, mathematically constructed in
ref.\cite{coaxial}, which propagate down a coaxial guide (in the
TM case): This picture is just taken from ref.\cite{coaxial}, \
but analogous X-pulses exist (with infinite or finite total
energy) for propagation along a cylindrical, normal-sized metallic
waveguide]. }
\end{center}
\label{fig27}
\end{figure}

experimentally produced by Saari et al.\cite{Saari97} in 1997 at
Tartu by visible light (Fig.\ref{fig26}), and more recently by
Mugnai, Ranfagni and Ruggeri at Florence by
microwaves\cite{Ranfagni}. \ In the theoretical sector the
activity has been not less intense, in order to build up ---for
example--- analogous new solutions with finite total energy or
more suitable for high frequencies, on one hand, and localized
solutions Superluminally propagating even along a normal waveguide
(cf. Fig.\ref{fig27}), on another hand, and so on.

\

\h  Let us eventually recall the problem of producing an X-shaped
Superluminal wave like the one in Fig.12, but truncated
---of course-- in space and in time (by use of a finite antenna,
radiating for a finite time): in such a situation, the wave is
known to keep its localization and Superluminality only till a
certain depth of field [i.e., as long as they are fed by the waves
arriving (with speed $c$) from the antenna], decaying abruptly
afterwards.\cite{Durnin,MRH} \ Let us add that various authors, taking
account, e.g., of the time needed for fostering such Superluminal
waves, have concluded that these localized Superluminal pulses are
unable to transmit information faster than $c$.
 \ Many of these questions have been discussed in what
precedes; for further details, see the second of refs.\cite{PhysicaA}.

\h Anyway, the existence of the X-shaped Superluminal (or
Super-sonic) pulses seem to constitute, together, e.g., with the
Superluminality of evanescent waves, a confirmation of extended
SR: a theory\cite{Review} based on the ordinary postulates of SR
and that consequently does not appear to violate any of the
fundamental principles of physics. \ It is curious moreover,  that
one of the first applications of such X-waves (that takes
advantage of their propagation without deformation) has been
accomplished in the field of medicine, and precisely ---as we
know--- of ultrasound scanners\cite{LuBiomedical,LuImaging}; while
the most important applications of the (subluminal!) Frozen Waves
will very probably affect, once more, human health problems like
the cancerous ones.

\

\

%
%

\

{\bf Acknowledgements}\\

For useful discussions they are grateful, among the others, to
R.Bonifacio, M.Brambilla, R.Chiao, C.Cocca, C.Conti, A.Friberg,
G.Degli Antoni, F.Fontana, G.Kurizki, M.Mattiuzzi, P.Milonni,
P.Saari, A.Shaarawi, R.Ziolkowski, and particularly A.Loredo and
M.Tygel.

\

\

\end{document}